\documentclass[
    acmsmall,screen,nonacm,%review,anonymous,
    prologue,dvipsnames
]{acmart}

\usepackage[english]{babel}

\PassOptionsToPackage{showframe}{geometry}

\usepackage{amsmath}
\usepackage{graphicx}
\usepackage{tikz}
\usepackage{listings}
\lstdefinestyle{listingstyle}{
    breakatwhitespace=false, 
    tabsize=2,
    keywordstyle=\bfseries\color{black},
    commentstyle=\itshape\color{gray},
    basewidth = {.30em},
    stringstyle=\color{Sepia},
    basicstyle=\ttfamily\color{darkgray},
    showstringspaces=false,
    columns=flexible,
}
\PassOptionsToPackage{hidelinks}{hyperref}

\title{Crowdsourcing subjective annotations using pairwise comparisons reduces bias and error compared to the majority-vote method}

\author{Hasti Narimanzadeh}
\orcid{0000-0001-7856-7419}
\email{hasti.narimanzadeh@aalto.fi}
\affiliation{%
  \institution{Aalto University}
  \streetaddress{P.O.~Box 11000}
  \country{Finland}
  \postcode{FI-00076}
}

\author{Arash Badie-Modiri}
\orcid{0000-0002-2027-360X}
\email{arash.badiemodiri@aalto.fi}
\affiliation{%
  \institution{Aalto University}
  \streetaddress{P.O.~Box 11000}
  \country{Finland}
  \postcode{FI-00076}
}

\author{Iuliia Smirnova}
\email{iuliia.smirnova@helsinki.fi}
\affiliation{%
  \institution{University of Helsinki}
  \streetaddress{P.O.~Box 4 (Yliopistonkatu 3)}
  \country{Finland}
  \postcode{FI-00014}
}

\author{Ted Hsuan Yun Chen}
\orcid{0000-0002-3279-8710}
\email{ted.hsuanyun.chen@gmail.com}
\affiliation{%
  \institution{George Mason University}
  \streetaddress{4400 University Drive}
  \city{Fairfax}
  \state{Virginia}
  \country{USA}
  \postcode{22030}
}

\ccsdesc[500]{Information systems~Crowdsourcing}

\keywords{crowdsourcing, subjectivity, comparison method, majority-vote method}

\begin{abstract}
How to better reduce measurement variability and bias introduced by subjectivity in crowdsourced labelling remains an open question. We introduce a theoretical framework for understanding how random error and measurement bias enter into crowdsourced annotations of subjective constructs. We then propose a pipeline that combines pairwise comparison labelling with Elo scoring, and demonstrate that it outperforms the ubiquitous majority-voting method in reducing both types of measurement error. To assess the performance of the labelling approaches, we constructed an agent-based model of crowdsourced labelling that lets us introduce different types of subjectivity into the tasks. We find that under most conditions with task subjectivity, the comparison approach produced higher $f_1$ scores. Further, the comparison approach is less susceptible to inflating bias, which majority voting tends to do. To facilitate applications, we show with simulated and real-world data that the number of required random comparisons for the same classification accuracy scales log-linearly $O(N \log N)$ with the number of labelled items. We also implemented the Elo system as an open-source Python package.
\end{abstract}

\begin{document}
\maketitle

\section{Introduction}
Human labels are consistently seen as the gold standard for producing data used to train or validate models and systems. These tasks include assessing image characteristics \cite{ribeiro2011crowdsourcing, wallace2020sketchy}, labelling text generated by humans or machines \cite{aroyo2019crowdsourcing,kane2020nubia,marchenko2020improving,logeswaran2018content}, validating images produced by generative adversarial networks \cite{denton2015deep,salimans2016improved,borji2019pros}, doing last-mile validation of AI predictions \cite{mohanty2019second,mohanty20181}, and creating specifications for data set evaluation \cite{balayn2019designing}. Due to demands for increasingly larger data sets, a substantial portion of such subjective manual labelling tasks has shifted toward crowd work in recent years \citep{kittur2013future,porter2020enhancing}. However, despite their ubiquity in data collection and measurement tasks, extracting and aggregating subjective information from crowdsourced human coders remain open areas of research \cite{zheng2017truth, uma2021learning}.

In this paper we focus on a particular class of subjective crowdsourcing tasks where the target quantity is a \textit{subjective construct} characterised by inherent subjectivity \citep{salminen2018inter}, such as content toxicity \citep{aroyo2019crowdsourcing,balayn2019designing,salminen2018online}, as opposed to one where there can be objectively defined targets \citep{sumner2020crowdsourcing,jang2022decreasing}. \textit{How do we obtain a valid classification or rating on a set of items for some subjective construct from a group of subjective human coders?} Consider, for example, the problem of identifying or quantifying the level of toxicity in online content \cite{aroyo2019crowdsourcing,balayn2019designing,salminen2018online}. The to-be-fit models for toxicity detection work by being optimised to best predict the average coder's assessment of toxicity, making input data critical to their performance. It is therefore integral to recognise that human assessment of toxicity is at best intersubjective and will be affected by numerous ``hidden variables'', namely social, linguistic, and economic factors that are often not explicitly observable in the assessment. As we describe below, these unobserved subjective factors increase random error and can often also introduce factors that researchers reasonably wish to disassociate from the construct in question (e.g., for algorithmic fairness purposes), which can be understood as a type of measurement bias \citep{li2020towards,hube2019understanding,cabrera2014systematic}. Crowdsourced toxicity labels, for example, have unfortunately been found to be positively associated with terms related to LGBTQ and race topics \citep{dixon2018measuring}.

Prior efforts have been made to improve labelling through subjectivity-aware tasks, including enhancement of questionnaire and task design \citep{hube2019understanding,wu2021toward}, an example being the quality-control workflow by \citet{drapeau2016microtalk}, which asks crowd raters to justify their answers or re-evaluate their decisions when faced with counter-arguments; uncovering and utilising disagreement and their sources among raters \citep{kairam2016parting,chang2017revolt,salminen2021problem}; model- or aggregation-based bias correction methods that rely on rater task histories \citep{li2020towards,kovashka2013attribute,wallace2022debiased,tao2020label}; and devising methods to collect labels from unskilled crowd raters that are on par with or superior to those from raters with domain expertise \citep{snow2008cheap,hovy2014experiments,dumitrache2017crowdsourcing}. Just as important, others have made efforts to conceptually and empirically sort through the nature of subjectivity in human-labelling tasks, such as classifying tasks by their level of subjectivity \citep{salminen2018inter} or identifying the different sources of subjectivity \citep{dumitrache2015crowdsourcing,kairam2016parting}.

\begin{figure}[b]
    \centering
    \tikzstyle{arrow} = [thick,->,>=stealth]
    \tikzstyle{startstop} = [rectangle, rounded corners, minimum width=4cm, minimum height=1cm,text centered, draw=black, fill=gray!20]
    \begin{tikzpicture}[node distance=4.75cm, font = \sffamily]
        \node (sb1) [startstop] {Pairwise Comparison};
    \node (sb2) [startstop, right of=sb1] {Elo Scoring};
    \node (sb3) [startstop, right of=sb2] {Score Distribution};
    \draw [arrow] (sb1) -- (sb2);
    \draw [arrow] (sb2) -- (sb3);
    \end{tikzpicture}
    
    \tikzstyle{startstop} = [rectangle, rounded corners, minimum width=4cm, minimum height=1cm,text centered, draw=black, fill=gray!0]
    \begin{tikzpicture}[node distance=4.75cm, font = \sffamily]
    \node (sa1) [startstop] {Item-wise Labelling};
    \node (sa2) [startstop, right of=sa1] {Majority Vote};
    \node (sa3) [startstop, right of=sa2] {Binary Labels};
    \draw [arrow] (sa1) -- (sa2);
    \draw [arrow] (sa2) -- (sa3);
    \end{tikzpicture}

    \caption{Different labelling approaches to obtaining measures of subjective constructs. The top row, highlighted in gray, shows our proposed pairwise comparison approach, which yields a continuous distribution of scores (e.g.~level of toxicity). The bottom row shows a traditional approach based on binary item-wise labels (e.g.~toxic or non-toxic).}
    \label{fig:pipeline}
\end{figure}
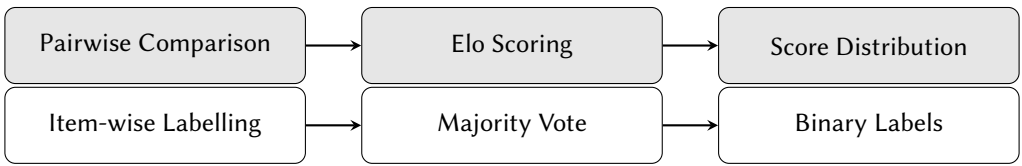

We add to this body of work by proposing a combined task design and aggregation approach for crowdsourcing measures of subjective constructs. Specifically, we describe a pairwise comparison-based labelling task, shown in the top row of \autoref{fig:pipeline}, where raters are given tasks of comparing the strength of the construct in question between two items, which we aggregate using the Elo rating system. We compare the pairwise comparison approach to the classic ``majority-vote'' labelling method, a ubiquitous mechanism where the same item is assessed by a specific number of people and the label for each item is selected based on the decision of a majority of voters. We are not the first to propose relative assessments for crowdsourced labelling \citep{parikh2011relative,chen2013pairwise,cheng2015flock,zou2015crowdsourcing,sunahase2017pairwise}, which we summarise below, but we do so explicitly in the context of obtaining subjective constructs, and combine the comparisons with a straightforward aggregation method, an open-source implementation of which is provided with this paper.

Further, we contribute to the definitional clarity of ``subjectivity'' by presenting a conceptual framework for crowdsourcing subjective constructs that maps this type of crowdsourcing task onto concepts from survey methodology, an idea that has been proposed elsewhere \citep{alonso2015debugging}. This framework lets us formalise what ``subjectivity'' means in measurement tasks, and show how subjectivity increases random error and measurement bias in aggregated labels. Namely, subjectivity increases variance and therefore increases random error, and coupled with necessarily imperfect task definitions means there are likely going to be sources of bias. We show that our pairwise comparison method reduces random error, as measured by binary-label $f_1$ scores. At the same time, while our method does not completely mitigate the issue of bias (which should be addressed in other steps of the rating pipeline \citep{li2020towards}), it outperforms the majority-vote method which actually inflates the inherent bias in the system.
 
Our work contributes to a wide range of literature that engages with the question of crowdsourcing and how to measure subjective constructs \citep{kairam2016parting, wallace2022debiased,salminen2018inter,dumitrache2015crowdsourcing,hube2019understanding,wu2021toward}. In addition to addressing the topic at a theoretical-level, mapping the task to a survey methodology problem, our work has clear implications for applied research. To facilitate applied work, we provide a reference implementation of the Elo rating system for pairwise comparisons as an open-source Python package under the MIT license.\footnote{Available at \anon[Redacted URL]{\url{https://github.com/hastinarimanzadeh/elo-rating}}.}

Our paper proceeds as follows. Beginning with Sec.~\ref{sec:theory}, we define our task of obtaining a valid subjective construct and formally describe the behavioural model of crowdsourced labelling. Here, we discuss how subjectivity enters the model, and how it introduces measurement error. In Sec.~\ref{sec:compare-v-vote}, we draw on previous work about comparisons and voting to discuss the advantages of comparison approaches over voting-based ones, both in terms of construct validity and implementation cost. Next, in Sec.~\ref{sec:methods}, we describe our proposed approach and our experimental setup. In Sec.~\ref{sec:results}, we present our results. We show that (1) for a wide range of subjective problems, the comparison method performs better than majority voting and the gap often gets wider the more subjective a task becomes; (2) the comparison method is more resilient towards spam, compared to the majority-vote method; and (3) the number of required comparison grows log-linearly with the number of items when used with the Elo rating system, not quadratically.

\section{Crowdsourcing subjective constructs}\label{sec:theory}
Building on previous research, we understand crowdsourcing measures for subjective constructs as a descriptive inference task of estimating the distribution of \emph{perceptions toward how items relate to a researcher-defined construct}. Such answer distributions can more accurately represent how people perceive the construct under study, and so the quest for devising crowdsourcing tasks conducive to estimating the population distribution as accurately as possible is an active area of research \citep{dumitrache2018capturing, chung2019efficient}. 

Exactly which aspects of the distribution to use in downstream research is an open question that touches on the existence and the nature of ground truths in subjective constructs \citep{aroyo2015truth,zheng2017truth}. A longstanding debate in the legal community surrounding how to define subjective constructs is instructive here. Namely, defining ``obscenity'' has historically been a challenge for the U.S. Courts system. It is a highly subjective term with immense legal implications \citep{goldberg2010two}. Going back as far as \textit{Roth v. United States} in 1957, the U.S. Supreme Court has constructed the hypothetical judge of obscenity to be the ``average person, [who is] applying contemporary community standards''. U.S. Supreme Court Justice Potter Stewart's famous 1964 statement ``I know it when I see it'' about defining ``hard-core pornography'' was in reference to \textit{Roth}, and was used then, and elsewhere \citep{sobieraj2011incivility}, to justify a data-driven and inductive way of defining subjective constructs. Indeed, when working with purely subjective constructs, as opposed to extraction of factual information \citep{sumner2020crowdsourcing} or skill-based assessments \citep{jang2022decreasing}, the crowd from which we derive the average judgement serves not as aggregated intelligence but instead as a reasonable \textit{definition} of the inter-subjective ground truth.

With this in mind, we proceed in this study with the premise that the population distribution surrounding perceptions of how items relate to the subjective construct of interest is usefully modelled as a unimodal distribution whose mean is meaningful. Specifically, we will treat the population mean as the ``ground truth'' and the target of the crowdsourcing task.\footnote{We recognise recent work exploring the nature of the ``ground truths'' and how they relate to the potential multimodality of crowdsourced distributions for measures of subjective constructs \citep{aroyo2015truth, zheng2017truth}. However, the population mean remains an important and often useful target in applied systems \citep{alonso2015debugging}.} This problem definition sets our study apart from others that focus on using crowd intelligence to complete tasks whose outcomes can be evaluated objectively \citep[e.g.][]{jang2022decreasing,sumner2020crowdsourcing}.

Following this logic, crowdsourcing subjective constructs and the challenges associated with this can be understood in terms of concepts from survey methodology. Specifically, our target is the population mean and we sample individuals from the population to estimate this quantity. Measurement error (i.e., deviations between the quantity inferred from the sample and the target measure) is either due to random error from sampling variation -- which is amplified in crowdsourcing tasks since usually only a very small number of responses are obtained -- or a type of response bias stemming from mismatch between rater responses and the target construct, which includes those due to under-defined tasks or unfair perceptions held by the crowd raters.\footnote{Sampling bias due to the unrepresentativeness of crowdsourced items is out of the scope of our study.}

``Subjectivity'', broadly defined as variations in how constructs are understood, fits into this framework as it contributes to measurement error by increasing the population variance surrounding the construct, thereby increasing random error, or by introducing unwanted components to the construct, thereby adding response bias. We discuss these in detail below.

\subsection{Model of crowdsourced labelling and sources of subjectivity}\label{sec:subjectivity-model}
In subjective crowdsourced tasks, definite answers can be elusive, not to mention susceptible to sources of inter-rater disagreement. Linguistic philosophers have long recognised different sources of ambiguity in meaning stemming from what corresponds in the data-labelling context to the ambiguity of data, differences in raters' perspectives, and ambiguity in annotation labels \citep{ogden1925meaning,knowlton1966definition}. These concepts have recently been adapted to research on subjectivity in crowdsourcing \citep{dumitrache2015crowdsourcing,aroyo2015truth,kairam2016parting}. \citet{kairam2016parting}, for example, identified various sources of disagreement among sub-groups of crowd raters in their entity annotation tasks of Twitter and Wikipedia data. Drawing from these concepts, we mathematically formalise our conception of subjectivity in crowdsourcing tasks below.

Following the previous discussion, we devised a simplified mathematical model of human perception and labelling of subjective constructs. We begin with an observable item, such as an internet comment, and a human rater, whom we ask to rate, for example, the item's level of toxicity. Intuitively, the rater will subconsciously (1) observe all aspects of the item, (2) give a toxicity component score to each observable aspect, (3) sum the component scores for a total item toxicity score, and (4) assign the item a toxicity label or numeric score, or compare it to another item.

\subsubsection{Definitional subjectivity}\label{subsubsec:definitional-subjectivity}
\begin{figure}
    \centering
    \includegraphics[width=\textwidth]{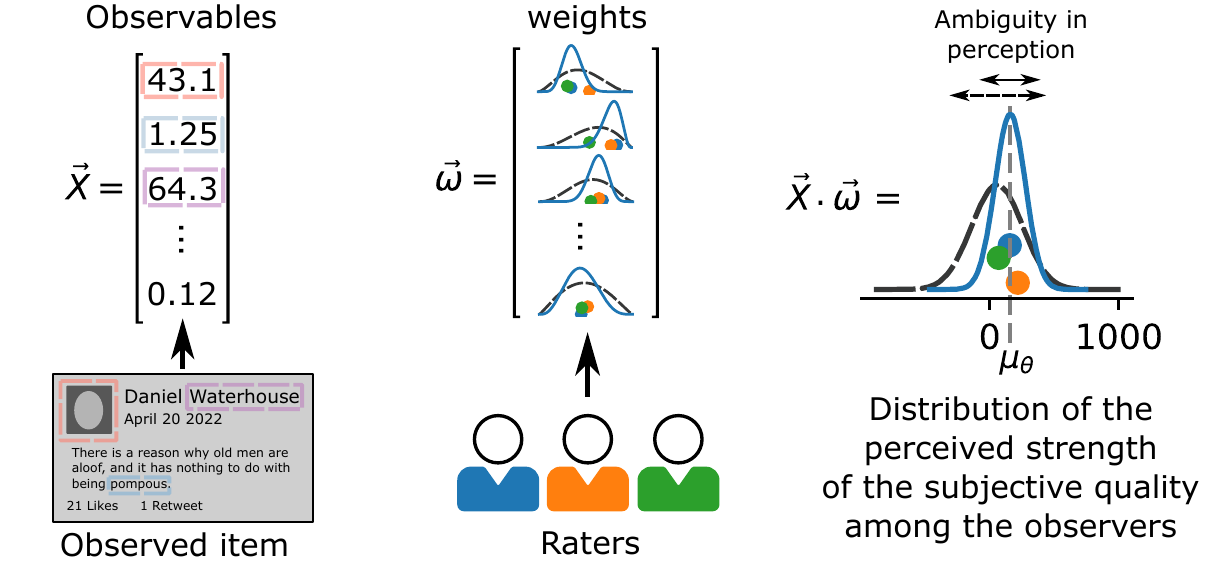}
    \caption{A simplified mathematical model of human understanding of subjective constructs. Let us assume vector $\vec{X}$ encodes all observable features of an item. For each rater there exists a vector of weights $\vec{\omega}$, normalised so that $\lVert \vec{\omega} \rVert = 1$, where $\omega_k$ determines the contribution of the observable $X_k$ in the strength of the subjective construct perceived by that rater $\theta_i$. The distribution of weights $\omega_k$ in the population of raters determines the overall ambiguity of perception for that item in the population. A more diverse set of opinions in the population, visualised here as the more widely distributed weights with dotted lines, produces more perceived strengths that are more varied across the population of raters. Conversely, more agreement in the weights, shown using solid blue lines, will result in higher overall agreement. In this paper, the standard deviation of the distribution of perceived strengths of the subjective construct is denoted as \textit{ambiguity in perception}.}
    \label{fig:framework}
\end{figure}
Let us encode all the objective observable facts about the item (e.g.~number of times profanity was used) as a very large vector of numbers $\vec{X}$. The elements of this vector independently encode all available information about the item.\footnote{Dependence between elements $a$ and $b$ can be account for by adding in an additional element capturing the interaction term $a \times b$.} In this simple model we assume that each element of $\vec{X}$'s contribution to the subjective construct is weighed by the rater according to a vector of weights $\vec{\omega}$, so that
\begin{equation}\label{eq:theta}
    \theta = \vec{X}\cdot\vec{\omega} = \sum_k X_k \omega_k\,,
\end{equation}
with the vector of weights being (arbitrarily) of unit length $\sum_k \omega_k^2 = 1$. The vector $\vec{\omega}$ is specific to each rater and is affected by their personal idiosyncratic preferences. Essentially, $\vec{\omega}$ is the rater's definitional understanding of the construct, and subjectivity exists because it differs across raters. The resulting value of $\theta$ denotes a rater's perception of the strength of the subjective construct in question. Figure~\ref{fig:framework} shows a schematic visualisation of this model. As defined above, our goal is to devise a method of estimation that recovers $\mu_\theta$, i.e.,~the population mean perceived strength of the subjective construct.\footnote{It is not possible to know without extensive study the distribution of $\omega_k$ in the population. We can, however, reasonably assume, based on generalisations of the Central Limit Theorem, that for an item with a large enough number of observables, the value $\theta_i$ of the same item across the rater population is sampled from a normal distribution $\theta \sim \mathcal{N}(\mu_\theta, \sigma_\theta^2)$. }

We define subjectivity as \textit{variation across individuals' conceptualisation of a given construct} (i.e., understanding and judgement) based on individual idiosyncrasies. If there is any variation in the population distribution of this conceptualisation that is not attributable to factors external to these individuals, we say there is subjectivity surrounding the construct and therefore the rated item.

Subjectivity factors into labelling tasks in multiple ways. At the definitional stage, which we just described, variations in preferences and experiences across individuals influence how much weight ($\vec{\omega}$) different things (i.e., observables in $\vec{X}$) are given when measuring a construct \citep{aroyo2019crowdsourcing,hube2019understanding}. \textit{Does profanity matter at all to text being perceived as toxic? If so, how much?} This kind of variation in what ``counts'' is a major source of subjectivity in labelling tasks. This type of subjectivity can be mitigated by clear coding rules \citep{aroyo2019crowdsourcing} -- \textit{do not consider the speaker's race when assessing the toxicity of what they said} --  but given the effectively infinite feature space of plausible observables in $\vec{X}$, subjectivity cannot be eliminated. Further, because many of these internal assessments are deeply-rooted biases, explicit prompts about biases can still fail to mitigate them \citep{hube2019understanding}.

As subjectivity about a given construct increases, the population variance of each $\omega_{k}$ increases, which is then reflected in increased variance in the distribution of $\theta$. The standard deviation $\sigma_\theta$ therefore can be understood as the \textit{ambiguity in perception}, which stems from the definitional process surrounding a subjective construct. Larger values of ambiguity in perception imply there will be higher variation in the perceived strength of the subjective construct among the population for a given item. As we illustrate in Fig.~\ref{fig:framework}, the consequence of higher ambiguity in perception is that there will be higher measurement error in the form of random error stemming from sampling variation, as the greater the population variance, the more likely a given sample mean will fall from the population mean. Standard crowdsourcing practices exacerbate this type of measurement error because samples (i.e.~the number of raters per item) tend to be small.

\subsubsection{Translational subjectivity}\label{subsubsec:trans-subjectivity}
Another source of subjectivity comes from raters' mental model when translating the latent $\theta$ into a concretely expressed measure, which can be through a classification task -- \textit{this comment is toxic} -- or a comparison -- \textit{comment A is less toxic than comment B}. Specifically, $\theta_i$ exists internally for each individual, and the act of encoding it (e.g.~threshold-based classification, quantitative rating, comparison) is also subject to variations across individuals.

If the rater is asked to make a binary classification, they compare their perceived strength of the subjective construct $\theta_i$ to some personal threshold $\theta_i^\text{threshold}$. As with the weight vector $\vec{\omega}$, the threshold is also specific to each rater and parameterise the sensitivity of the rater to the subjective construct. We refer to this type of subjectivity as \textit{diversity in threshold}.

On the other hand, if the rater is asked to make a comparative assessment of two items, that is, to determine which of two items is stronger on the subjective construct, the rater will perceive the magnitude of the construct in the two items, say $\theta_1$ and $\theta_2$, and then compare them. Based on prior work on rater agreement, it is reasonable to assume that it is easier for a rater to compare two items that are perceived to be more different to each other with respect to the strength of the subjective construct in question \citep{salminen2019online}, i.e., when $|\theta_1 - \theta_2|$ is larger. In our model, the rater is not able to make a reliable comparison if the difference in perceived strength of the two items $|\theta_1 - \theta_2|$ is smaller than some cutoff value $\delta_\theta$. We refer to this parameter $\delta_\theta$ as the \textit{ambiguity in comparison}. 

\subsubsection{Subjectivity-induced bias}\label{subsubsec:bias}
\begin{figure}
    \centering
    \includegraphics[width=\textwidth]{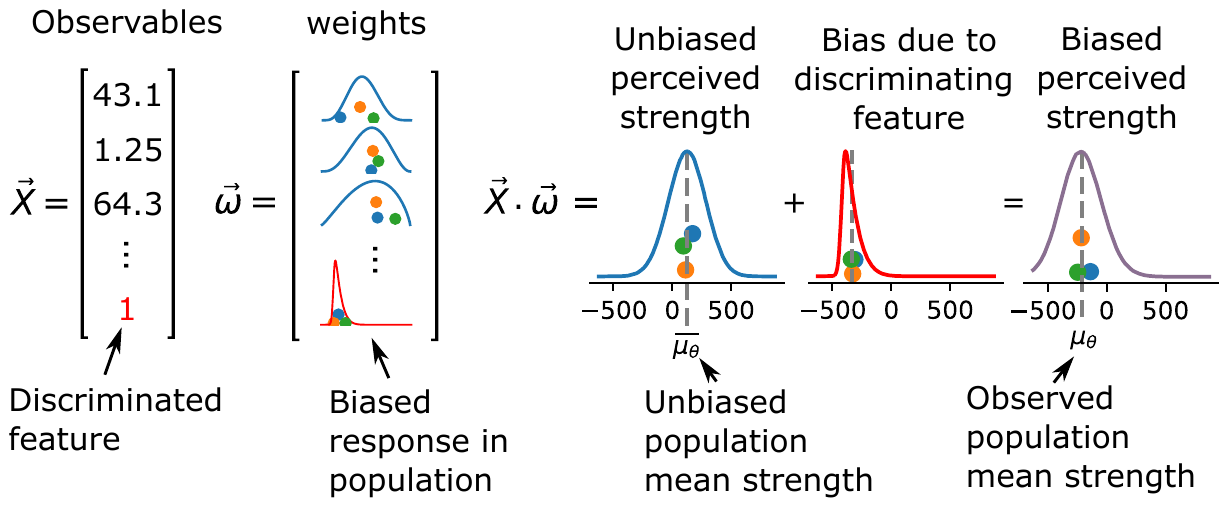}
    \caption{A simple mathematical model of bias induced by the existence of a discriminatory feature in an item. Some items might display features that the researcher decides, for ethical or scientific reasons, should not affect the perceived strength of the subjective construct. Existence of LGBTQ and race-based terms contributing to toxicity labels exemplifies this issue \citep{dixon2018measuring}. In the simplified model of human understanding of subjective constructs presented here, these features induce an error (and no bias) if the distribution of their associated weights has a mean of zero. On the other hand, if the discriminated feature has a weight distribution with a mean much larger or smaller than zero, it induces a bias in the perceived strength of the subjective construct in all items that exhibit the feature. Here, we see that a heavily-skewed distribution of weights (in red) associated with the discriminated feature significantly affects the perceived strength of the subjective construct for the items exhibiting that feature.}
    \label{fig:bias}
\end{figure}

Finally, we discuss the special case where definitional subjectivity, discussed above, becomes a source of measurement bias. Specifically, instead of targeting the population mean $\mu_\theta$ (i.e.~where the researcher recognises the existence of definitional subjectivity but is agnostic about what ``counts'' in the construct's definition), there are instances where researchers would wish to exclude certain features from the definition of a construct and therefore not have them factor into the crowdsourced measure. In these cases, if researchers do not or cannot fully define the task, which we established is difficult \citep{hube2019understanding}, the presence of subjectivity among the raters about whether these features should be a part of the construct's definition means $\mu_\theta$ becomes a biased version of the target construct, which we illustrate in Fig.~\ref{fig:bias} with a schematic visualisation. In our model, this scenario is presented as researchers desiring an element in $\vec{\omega}$ to be zero when its population mean is in fact nonzero.

What we have described here is effectively a form of response bias stemming from the inability of researchers to control unwanted definitional subjectivity. This scenario is becoming increasingly common as researchers start to recognise the importance of excluding unfair or otherwise undesirable observables from entering their labelled data, and in turn their models or algorithms. For example, researchers classifying internet comments into toxic and non-toxic categories are likely to want the labels to be unaffected by ethnic or racial features present in the profile image of the poster.

\section{Comparison versus majority-vote methods}\label{sec:compare-v-vote}
Having now defined our target (i.e.~a subjective construct measured as its population mean $\mu_\theta$) and presented the associated challenges stemming from different sources of subjectivity, we move to discuss the relative benefits of two widely used ways to formulate crowdsourcing tasks, the comparison and the majority-vote methods which we presented in Fig.~\ref{fig:pipeline}. Drawing from previous literature, we discuss how the two methods compare in terms of the validity of the obtained measure when they are faced with different sources of subjectivity. We also review claims about their costs.

\subsection{Validity}
The two methods we study in this paper have been previously compared elsewhere in the literature. One line of research focuses on the cognitive benefits of pairwise comparison tasks over item-wise labelling tasks at the individual-level, while another considers how these tasks compare specifically in crowdsourcing contexts where information gathering tends to favour breadth over depth.

\subsubsection{Cognition-based benefits of comparative over absolute judgements}
From the psychology and cognition literature, \citet{goffin2011all} argue that when individuals assess performances, comparative judgements are superior to absolute judgements, which are the building block of the majority-vote method, because the latter often lacks a clear anchor. On the other hand, anchoring assessments to an external comparison point provides respondents ``shared reference points on which to base their judgements'', which ``may reduce error variance by increasing respondents' agreement about the meaning of scale points \citep[][p.53]{goffin2011all}''. \citet{gentner1997structure} find a similar mechanism, that explicit juxtapositions highlight similarities and differences, which provides clarity on the compared construct. This concept has been usefully applied to human-labelling contexts, such as using comparison-based crowdsourcing to extract features \citep{cheng2015flock,zou2015crowdsourcing}.

\subsubsection{Removing diversity in threshold through relative anchoring}
Other work from human-labelling fields, such as image assessment \citep{parikh2011relative} and toxicity labelling \citep{aroyo2019crowdsourcing}, focus explicitly on the benefits of \emph{relative anchoring}, where to-be-labelled items are assessed against each other, dispensing with the need for either individuals' internal anchors or external anchors. In their seminal piece on relative assessments for image labelling, \citet{parikh2011relative} offer assessing smiling as an illustrative example of the relative anchoring concept. Given three images of (1) someone clearly smiling, (2) someone clearly not smiling, and (3) someone having a borderline expression between smiling and non-smiling, it is difficult for absolute ratings to place the third category, but the comparative rating can always confidently distinguish between any pair of images.

This argument about relative anchoring directly maps onto our discussion of translational subjectivity from Sec.~\ref{subsubsec:trans-subjectivity}. Specifically, because compared items serve as reference points to each other, the comparison method is able to remove diversity between raters' thresholds, one of our identified sources of subjectivity, from the labelling procedure.

\subsubsection{Avoiding measurement error inflation for high-subjectivity constructs}\label{subsubsec:avoid-error}
Another line of work focuses on how the majority-vote method obscures the amount of uncertainty surrounding specific items \citep{salminen2021problem,aroyo2019crowdsourcing}. By construction, the majority-vote uses popularity to adjudicate between two side of a binary label, thereby forcing agreement. However, the very fact that the majority-vote needs to be employed in the first place implies there is variation in the population surrounding the subjective construct being measured, which can be shown through inter-rater agreement scores \citep{salminen2021problem,salminen2018online,alonso2015debugging}.

The discarded labels are not random, but are instead due to the raters' subjectivity, which as we argued can stem from both definitional and translational reasons. Further, the discarded data will always be from minority raters holding less popular opinions. The majority-vote method, then, moves the estimated mean $\hat{\mu_\theta}$ further from the true population mean by non-randomly discarding up to [$\frac{1}{3}$,$\frac{1}{2}$) of the ratings. As measurement error due to population variability in crowdsourcing tasks is already likely to be high because only a small sample size is used for each item, this kind of data removal will further exacerbate the problem. Taken to the extreme, the majority-vote method is equivalent to running a survey and claiming that the population can be characterised by the modal response category.

Instead, as \citet{aroyo2019crowdsourcing} argue, for inherently subjective tasks (e.g., measuring subjective constructs as we do), a comparative ranking would result in better overall agreement between raters compared to an item-wise absolute rating due to the fact that the comparison method can capture the diversity of opinions between raters as opposed to optimising for consensus. This is achieved when using a pairwise comparison approach, as it does not discard any data. In fact, more than not wasting data, because the comparison approach allows each item to be observed multiple times but as part of different unique item pairs, it provides more information for aggregation algorithms such as the Elo method we describe in Sec.~\ref{eq:elo} to produce a more accurate estimate of $\theta$'s population distribution.

\subsubsection{Avoiding bias and spam inflation}\label{subsubsec:avoid-bias}
Finally, following the immediately preceding discussion from Sec.~\ref{subsubsec:avoid-error}, we hypothesise that comparisons will outperform majority-voting over item-wise labels when dealing with the kind of subjectivity-induced bias we discussed in Sec.~\ref{subsubsec:bias}. We start by recognising that should this type of bias exist, it will affect both methods. By Eq.~\eqref{eq:theta}, definitional subjectivity (captured by the presence of unwanted elements with nonzero mean in $\Vec{\omega}$) is a part of $\theta$, which means that because both the comparison and majority-vote methods target the population mean $\mu_\theta$, neither method do anything to \textit{remove} the bias. 

However, as the majority-vote discards observations based on popularity, it can yield highly biased estimates for some items due to random sampling, which we speculate will \textit{inflate} the extant population bias.\footnote{Depending on the distribution of weights $\omega_k$ for the bias-inducing element in $\Vec{\omega}$, this behaviour could plausibly reduce bias, but this is largely idiosyncratic to particular cases and therefore cannot be relied on as a feature of the method.} On the other hand, because pairwise comparisons are able to more accurately reflect the population distribution of $\theta$, it should also more accurately capture the unwanted bias component in $\mu_\theta$.

The superiority of the comparison method here, then, is in being able to avoid inflating measurement bias. A similar line of reasoning regarding the aggregation of unwanted information can be applied to the effect of spammers (i.e.,~raters who complete tasks randomly or semi-randomly), which is an active research area in crowdsourcing \cite{difallah2012mechanical,vuurens2011much}. 

\subsection{Cost}\label{subsec:cost}

Despite all the ostensible benefits of using comparisons, costs, specifically the cost of performing all possible number of pairwise comparisons grows quadratically with the number of items being rated. This serves to be an impediment to adopting the comparison method over the majority-vote method, whose total possible number of tasks grows linearly with the number of items, especially when dealing with large data sets \cite{guo2018experimental}. Further, previous studies have hypothesised that a certain level of labelling quality requires many more comparison tasks or an algorithmically selected set of comparison tasks, compared to the majority-vote tasks \cite{jang2022decreasing, guo2018experimental}. \citet{ye2013combining}, for instance, propose a method of augmenting the comparison method with absolute judgement to improve the labelling quality. They also propose an active method for determining the next batch of comparisons and absolute judgement tests based on previous batches of results. Similarly, \citet{jang2022decreasing} and \citet{guo2018experimental} propose an active method for determining future comparisons to reduce the number of comparisons required to arrive at a full ordinal ranking. 

In this paper, we show that depending on the subjectivity of the problem a comparable number of comparison tasks can provide as good, or better quality of labels compared to the majority-vote tasks, and that for a randomly selected set of comparisons combined with a simple implementation of the Elo rating system, the number of required comparisons for a constant level of labelling quality scales log-linearly, not quadratically, with the number of items.

Finally, we note here that researchers have argued that comparisons are often easier and faster to make compared to the majority-vote method when assessing the relevance of search results in information retrieval systems using crowd-sourcing \citep{carterette2008here}. A faster and easier task very commonly means that a higher number pairwise comparison tasks can be performed with the same amount of time and effort spent by the raters compared to tasks that rely on an absolute single-item judgement. 

\section{Methods}\label{sec:methods}
In the remainder of this paper, we explore questions raised in the preceding discussion regarding the relative strengths of the majority-vote and pairwise comparison methods when crowdsourcing measures for subjective constructs. We compare each method's response to the different types of subjectivity we discussed in our theoretical framework, including both definitional and translational sources of subjectivity. We also assess their performance in dealing with the presence of subjectivity-induced bias as we hypothesised in Sec.~\ref{subsubsec:avoid-bias}. Finally, we study the scaling behaviour of the pairwise comparison method to establish how the number of required comparisons to keep the same level of accuracy changes as the number of items to be rated grows.

Before moving to our results, we present the methods we used. First, in Sec.~\ref{sec:elo}, we introduce the label aggregation approach we adopt in this paper, namely the Elo rating system. Next, we describe in Sec.~\ref{sec:simulation} how we quantify our subjectivity formalisation through the simulation of items and raters. Finally, in Sec.~\ref{sec:experiments}, we summarise our experimental pilot of Twitter conversations which we use to corroborate our simulated $O(N \log N)$ scaling of random comparisons.

The code required for reproducing the numerical results presented in this paper, along with all the necessary instructions and data, are available as a Git repository at \anon[Redacted Git Repository URL]{\url{https://github.com/hastinarimanzadeh/elo-paper-reproduction}}.

\subsection{The Elo rating system}\label{sec:elo}
As we showed in Fig.~\ref{fig:pipeline}, the results from pairwise comparisons undergo an algorithmic transformation before ending up as ratings, which can then be binarized by rank for classification purposes.
The problem of assigning ratings to the strength of each player based on their performance in one-on-one matches against others appears frequently in games and sports. Historically, hand-crafted rating systems relied on an often intrinsic judgement of each player's merit or achievements in each game, combined with a myriad of statistical and heuristic methods specific to the game in question. In this paper, we dispenses with these historical approaches and propose using the Elo rating system as a simple yet effective method of aggregating the results of comparisons performed by crowdsourcing raters. 

The Elo system works by relying on statistical estimation of ratings solely based on the final outcome of performed matches. In short, for each match, the Elo system makes a prediction of the outcome probabilities in a zero-sum games based on the two participating players' difference in rating. The two players \textit{wager} some fixed amount $k$ of their rating based on the predicted probabilities. If the outcome of the match is closer to the predicted outcome (e.g., a much stronger player wins against a much weaker player) fewer rating points change hands. In the case of an unlikely outcome, a larger share of wagered points is transferred to the winning player.

The predicted outcome of a match depends on the difference in the ratings of the two players. The system assumes that the actual game performance of a player in a single game is a random variable, often assumed to be normally or logistically distributed, with a mean equal to the player's true rating (representing their aptitude or strength) and a scale arbitrarily set. The Elo system makes the simplifying assumption that all players have the same scale for their \textit{game performance distribution}. A single match between two players can be simulated by each player drawing a number from their respective game performance distribution. The probability of a positive outcome for each player is equal to the probability of their drawn number having a larger value.

Many real-world rankings, such as ranking of countries by population and performance of sports players, show a heavy-tail decay of scores as ranks increase \cite{iniguez2022dynamics}. At first glance, it might seem that the Elo system assumes a specific distribution for the game performance. Instead, the system makes an assumption about the \textit{distribution of relative performance of two players} (i.e.~their respective probability of winning) given their difference in rating. In other words, the inherent assumption of this method is of a normal performance distribution, often called the Thurstone--Mosteller model, or based on a logistic performance distribution referred to as Bradley--Terry model. Though it has been shown that in a practical setting the two models produce virtually equivalent results \cite{stern1992all} the latter model, based on logistic distribution, is commonly used for ranking human competitive behaviour such as rankings for chess players and in online multi-player games, possibly due to its nicer analytical behaviour \cite{menke2008bradley}.

A commonly used formulation, based on the logistic distribution assumption is as follows:
\begin{equation}\label{eq:elo}
    E_A = \frac{1}{1 + 10^{\frac{R_A - R_B}{400}}}\,,
\end{equation}
where $R_A$ and $R_B$ indicates ratings (aptitude) of the players $A$ and $B$. $E_A$ indicates the probability of a positive outcome of the game for player $A$. As possible outcomes range from 0 to 1 the expected outcome for each player is the same as the probability of a positive outcome for that player. The use of base 10 and denominator 400 makes it easier to mentally compare rating values, as a difference of 400 rating units is equivalent to a ten times larger expected score.

After a single game, the ratings are adjusted based on the difference in the realised outcome compared to the expected outcome of the game
\begin{equation}
    R^{\prime}_A = R_A + k(S_A - E_A)\,,
\end{equation}
where $S_A$ is the realised outcome of the match for player $A$. The rating for player $B$ is also adjusted using the same formula. The parameter $k$ is a tuning parameter, roughly similar to the learning rate parameter in many machine learning methods, which indicates the maximum possible point adjustment per match. A high value for $k$ results in larger adjustments to the ratings after each match, at the cost of higher overall noise. Conversely, selecting a lower value of $k$ requires more matches to arrive at equilibrium in ratings.

In addition, we found it often helpful in practice to run the rating system with multiple cycles on the available set of comparisons, generally similar to the deep learning common practice of cycling learning data in multiple epochs. At each epoch, the set of all available comparisons are used in random order for updating the ratings. This can be repeated until an equilibrium is reached. 

The Elo rating system has nice properties for our application. 
While the zero-sum nature of the Elo rating system might at first glance appear a weakness, we argue that this fits well with our goal of labelling items based on the population average. As results of each comparison changes the rating of two items by equal amounts but in opposing directions, the mean ratings value of all items will always stay the same. If all items start with an initial rating of zero, the label of each item can be determined by simply checking the sign of its estimated rating. The rating is directly related to the probability of that item being selected in a comparison with a hypothetical item that possesses the exact average of the intended subjective construct compared to all other items as construed by the population of the raters, as can be surmised from Eq.~\eqref{eq:elo}.

We also elected to use the Elo rating system in part due to its simplicity. In the next two sections we will show both empirically and using a simple mathematical model that despite relying on the most simple approach of randomly comparing items and aggregating the results using the Elo rating system, we still arrive at a scalable and robust estimation of popular average labelling of subjective constructs. 

An open-source reference implementation of the Elo rating system is provided as a Python package in the official Python Package Index. The package can be installed using the command \anon[Redacted pip install command]{\texttt{pip install elo-rating}} with the documentation available at \anon[Redacted documentation URL]{\url{https://github.com/hastinarimanzadeh/elo-rating}}. The package provides an Elo class that computes ratings and rankings based on a list of provided results of comparisons. Further details on the interface are delineated in the documentation.

\subsection{Simulation}\label{sec:simulation}
The relative performance of the comparison and majority-vote methods when faced with different sources of subjectivity can be studied using a simple agent-based model of the problem. We present here such a model, which encompasses the subjectivity model we previously described in Sec.~\ref{sec:subjectivity-model} and schematically illustrated in Fig.~\ref{fig:framework}. We follow this with details on the comparison and majority-vote methods simulation.

\subsubsection{Setup}
Let us assume $N$ items, e.g., internet comments, each with a ``true'' rating, indicating the inherent value of the construct in question as it would be judged by the average rater for each of the items, e.g., toxicity. For the purposes of this simulation, we draw the true ratings from a normal distribution with an arbitrary mean of 0 and a standard deviation of 1. A perfect binary classification system would be able to distinguish all items with ratings above 0.

We also simulate $M$ raters, each with a personal threshold of what they would consider a significant enough presence of the intended construct, e.g., strong enough toxicity that impels them to recognise a certain comment as toxic. The personal thresholds are drawn from a normal distribution with the same mean as the mean true rating of items and standard deviation set by the parameter \textit{diversity in threshold}, directly mapping to the translational subjectivity we discussed in Sec.~\ref{subsubsec:trans-subjectivity}.

To contextualise the definitional subjectivity we discussed in Sec.~\ref{subsubsec:definitional-subjectivity}, for each item a rater observes a perceived rating for that item, drawn from a normal distribution with mean set to the true rating of the item and a standard deviation set by the parameter \textit{perception ambiguity}.
The perceived rating of one item for one rater is fixed across different observations by the same rater at least in the time scale of one study, as it is assumed to be a function of past experiences, internal priorities, and their definitions of the construct in question. A rater can classify one item by comparing their perceived rating of that item with their \textit{personal threshold}.

\begin{table}[htb]
    \caption{The set of parameters used in the agent-based simulation setup.}\label{tab:parameters}
    \begin{tabular}{p{0.3\textwidth}p{0.6\textwidth}}\toprule
         \textbf{Parameter name} & \textbf{Description}  \\ \midrule
         Ambiguity in perception & Standard deviation of raters' perceptions $\theta$, defined in Sec.~\ref{subsubsec:definitional-subjectivity} \\ 
         Ambiguity in comparison & Minimum difference between perceived strength of two items ($|\theta_1 - \theta_2|$) below which the two items cannot be ordered by a rater, defined in Sec.~\ref{subsubsec:trans-subjectivity}\\
         Diversity in threshold & Standard deviation of raters' personal thresholds for the subjective construct, discussed in Sec.~\ref{subsubsec:trans-subjectivity}\\
         Rater bias & Rater's weight for a discriminatory feature in the observables vector, defined in Sec.~\ref{subsubsec:bias}\\
         Fraction of spammers & Fraction of spammers, raters who vote or compare randomly \\
         Number of items & Items to be labelled\\
         Number of comparison tasks & Pairwise comparisons among all items\\
         Number of votes per item in majority-vote & Unless stated otherwise, 3 votes per item\\ \bottomrule
    \end{tabular}
\end{table}

A rater can also be given a comparison task, for which they compare two items based on the perceived ratings of the items. It is, however, inherently more difficult to confidently compare items when the difference in the magnitude of the intended construct is small. We identified such an ambiguity to stem from translational subjectivity in Sec.~\ref{subsubsec:trans-subjectivity}. To allow for this in our model, if the difference in the perceived ratings of the two items is smaller than a certain threshold, as determined by the parameter \textit{comparison ambiguity}, the rater identifies them as equal. Conversely, a \textit{comparison ambiguity} of zero results in no rater selecting the equal option, but simply selecting whichever item that corresponds to a higher perceived rating for that rater.

Spam is an unfortunate reality in crowd-sourcing. For some problems, it is possible to systematically detect spammers based on various methods. \citet{kuang2020spam} for instance uses a heterogeneous network embedding approach to detecting spam raters based on defined collusion patterns. These methods become more difficult to implement for the case of subjective problems \citep{alonso2015debugging}. To assess the susceptibility of the two methods to spam, a fraction of raters can be simulated as spammers, essentially voting or comparing presented items at random regardless of the items' ratings or perceived ratings.

\subsubsection{Simulation}
For each item, three random raters are drawn from the pool of $M$ raters (which might include spammers) and each rater is asked to provide a binary classification of that item (a vote) based on comparing their perceived rating of that item and their personal threshold. The three votes are recorded and items are labelled using a majority-vote system. Next, a number of pairs of items are drawn from the pool of all possible combinations of size 2 of $N$ items for comparison. Each item pair is passed to a rater who compares them based on their perceived ratings and comparison ambiguity. A binary classification is established by running the Elo ranking algorithm on the comparison results and classifying items based on their final Elo ratings being larger than 0.

\subsubsection{Evaluation}
For each combination of parameters, the performances of the two classification methods are measured as compared to the ground-truth binary labels based on the true rating of each item. For both methods, the positive label $f_1$ score of each run is calculated. The mean difference in $f_1$ scores across an ensemble of 25 independent runs determines which binary classification method worked best in each specific scenario.

\subsection{Scaling experiment: Controversiality of climate change conversations on Twitter}\label{sec:experiments}
To confirm the scaling behaviour and the properties of the comparison method, we also performed a set of real-world experiments. In these experiments, the crowdsourced task was to classify a conversation on Twitter, consisting of the main tweet and its whole response tree, as either controversial or non-controversial. The goal of the experiment is to study the scaling behaviour of the comparison method and provide an example of how a small-scale pilot study can help estimate the number of required comparisons for a larger study.\footnote{As discussed in our problem definition (i.e.~Sec.~\ref{sec:theory}), when working with subjective constructs like the controversiality of a conversation, the ``ground truth'' for a given item is that item's population mean perception $\mu_\theta$. This target quantity can only be estimated, as even when working with a population census, which removes random error, labelling is still subject to the different types of subjectivity we outlined above. While it is possible to conduct a large-scale empirical study aimed at extracting the population distribution of each conversation as precisely as possible for the purpose of benchmarking different crowdsourced labelling approaches, we believe this is beyond the scale of our current study. We therefore focus our real-world experiment on studying scaling properties.}

To this end, we selected a sample of 41 conversations related to the topic of climate change. We sampled these conversations from all Twitter conversations between February 2019 and September 2021, where the original post contains at least one of the list of 13 keywords or 36 hashtags commonly associated with climate science, climate action, or climate scepticism, and the conversation involves at least three unique users.\footnote{Tweets containing phrases ``political climate'' and ``precip:'' (standing for precipitation, used in reporting of daily weather forecast) as well as retweets were excluded.}

We constructed a set of crowdsourcing tasks for all 820 possible comparisons between two given conversations. The comparison task stated the following prompt: ``Read the following two Twitter conversations. Which of the following two conversations shows a \textit{higher level of disagreement} among its participants?''. Each time a conversation was presented in a task, the replies were presented in random order and limited to a maximum of five first-level replies and their second-level replies. Each of the 820 possible comparisons was presented to two raters on Amazon Mechanical Turk. The raters had three possible choices for each comparison. They could select the conversation on the left, on the right, or judge them to have an equal level of disagreement. Each comparison was replicated across two different raters for a total of 1640 crowdsourcing tasks. Special care was given to make sure no single rater can perform more than 50 comparison tasks to simulate responses from a diverse population.

For the scaling experiment, we take the final ratings calculated using all 1640 task answers as a benchmark. To measure the change in accuracy as a function of the number of comparisons, we selected random subsets of the comparisons and fed them into the Elo system to arrive at ratings and predicted labels. Comparing these predicted labels to those calculated using the full suite of comparisons gives us an estimate of the accuracy. We also studied the effect of the number of items on the accuracy of classification by selecting samples of different sizes from the original 41 items, repeating the experiment and comparing the results.

\begin{table}[tb]
    \centering
    \caption{Climate twitter conversations across a range of controversiality. Replies to the top-level tweet are marked with arrows. Those highlighted in red ostensibly disagree with the top-level tweet. Higher rank means higher levels of controversiality based on Elo scoring of crowdsourced comparisons.}
    \begin{tabular}{p{0.02\textwidth} p{0.015\textwidth} p{0.85\textwidth}} \toprule
Rk. &  \multicolumn{2}{p{0.9\textwidth}}{Conversation} \\ \midrule

1 & \multicolumn{2}{p{0.9\textwidth}}{No, they're not.  Every study in 2019 and prior concludes their ineffectiveness.  Now that it's political they're suddenly effective. Science doesn't lie, but scientists with political agendas do.  As we have already learned from climate `scientists.'} \\ 
 & ${\color{red}\hookrightarrow}$ & The virus is spread through expectorant, meaning the stuff coming out of your nose and mouth when you sneeze and cough. That's what masks are meant to stop, and they're effective. \\
 & ${\color{red}\hookrightarrow}$ & Maria, it's true, masks are effective. \\
 & ${\color{red}\hookrightarrow}$ & How are you gonna do research on how effective are masks in 2019? It's 2020. This is covid 19. Something doctors have never studied before so everything you've supposedly "read" is not valid \\ \midrule
 
 20 & \multicolumn{2}{p{0.9\textwidth}}{Big News! Starbucks to ditch dairy for the environment. Starbucks admits that cow's milk is its biggest source of carbon dioxide emissions, yet it's still charging up to \$0.80 more for lattes made with soy, coconut, or almond milk rather than dairy.} \\ 
 & $\hookrightarrow$ & Hi I just have a query! I heard some nutritionist saying our body uses 80\% of energy in digesting food is it true? \\
 & $\hookrightarrow$ & That is really great news! I would say the best news of 2020. Hope is it for all countries. It will serve as role model for others to change! \\
 & ${\color{red}\hookrightarrow}$ & Almond milk is still pretty terrible for the environment... \\
 & $\hookrightarrow$ & If they're trying to promote environmental health, why charge more money for a latte? @Starbucks you can do better! \\
 & $\hookrightarrow$ & They have to bring in oat milk. Best for the environment and so good with coffee!! \\ \midrule
 
40 & \multicolumn{2}{p{0.9\textwidth}}{It's snowing in New York City today. After two years in a tropical climate and three in a desert, it's a sight for sore eyes.} \\
 & $\hookrightarrow$ & You weren't in Boise for Snowmageddon? \\ 
 & $\hookrightarrow$ & Just like being back in the mitten     \\ 
 & ${\color{red}\hookrightarrow}$ & Um, no... \\
 & $\hookrightarrow$ & I grew up in Jersey. I always loved the first snow. It's so beautiful. The lights shine on it and it looks like diamonds.  Stay warm Paul.\\ \bottomrule
    \end{tabular}
    \label{tab:convos}
\end{table}

Finally, while we do not conduct a large-scale survey to estimate the ground truth required for assessing the relative performance of the pairwise comparison and majority-voting methods, we provide confidence that our proposed approach yields reasonable results in the absolute by showing conversations and their estimated controversiality from our Twitter experiment. In Tab.~\ref{tab:convos}, we present three conversations across the high, moderate, and low ranges of controversiality estimated using our proposed approach. These three conversations are specifically chosen from their respective ranges because they are succinct enough to be presented in tabular form, but all the the experiment data, including the Tweet IDs, the crowdsourcing responses, and their Elo ratings are available in a Git repository at \anon[Redacted URL]{\url{https://github.com/hastinarimanzadeh/elo-paper-reproduction}}.% and as the example data directory for the Elo reference implementation package.

\subsubsection{Ethical considerations}
We have taken measures to ensure that we adhered to the necessary ethical considerations for running this crowdsourced experiment. While all tweets in the experimental data were public tweets, we anonymised the tweets such that no personally identifiable information from the posters was shown to the raters. When displaying each tweet, random usernames and random profile pictures were generated to replace the real ones and links and embedded images and videos in the tweets were redacted. Moreover, we did not gather any personal information from the raters. Finally, there was no experimenting or manipulation of information for the data presented to the raters and the entire process was minimally intrusive. The raters were paid 0.20\$US per comparison for a target average hourly wage of 12.00\$US. Based on our own pretests, we estimated each comparison to take approximately one minute to complete. Reviews from actual crowd raters on Turkerview ($n=8$) show a median completion time of 44.5 seconds which translates to an average hourly wage of 16.18\$US.

\section{Results}\label{sec:results}
In Sec.~\ref{res:subjectivity} we discuss how the various facets of subjectivity quantified in Sec.~\ref{sec:simulation} affect the accuracy of binary classification using the comparison and majority-vote methods, as expressed by their $f_1$ scores. As we show below, we find that the comparison method yields higher $f_1$ scores than the 3-vote majority-vote method either when there is an equal number of comparisons and votes, or has the room to scale with increasing resource relative to majority-voting. We find that the comparison method is more robust to spammers and sources of unwanted discriminatory bias.

In Sec.~\ref{res:scaling}, we proceed to simulate the scaling behaviour of the comparison method, computing and comparing its labelling accuracy as the number of items changes. We show that the number of required comparisons for a specific level of accuracy scales log-linearly with respect to the number of items to be labelled.

\subsection{Simulation results}\label{res:subjectivity}
In order to compare the classification performance of the majority-vote method to the combination of comparison method and Elo system, we simulated a set of $N = 512$ items with true ratings drawn from $\theta \sim \mathcal{N}(0, 1)$. Each item can be assigned a true binary label based on comparing their true ratings to the population mean $\theta > 0$. We also simulated a set of 100 raters, where each rater has a personal threshold drawn from $\theta_i^\text{threshold} \sim \mathcal{N}(0, 0.5)$, and their ambiguity in perception standard deviation and ambiguity in comparison are set to 0.5. The items are then classified with a majority-vote system with three votes by different raters, and subsequently with the comparison system presented in this paper, with $n_\text{comparisons}$ comparisons randomly selected sample from the pool of all possible comparisons. Each method then predicts a binary label for each item. The accuracy of the two methods is calculated in the form of the positive label $f_1$ score, where for every item the label predicted by each method is compared to its true label.

At each step, one of the parameters of the simulation (from the first five parameters in Tab.~\ref{tab:parameters}) is modified and the results are studied. The simulations for each parameter are repeated multiple times, and we compute the mean and the $5 \sigma$ confidence intervals of the values of interest (often the difference in $f_1$ scores of the two methods, $f_1^\text{comparison} - f_1^\text{majority}$). These results are summarised in the figures that follow.

For most of these figures, we plot the difference in $f_1$ scores either on the y-axis or the z-axis of a surface. Regions above zero indicate better performance by the comparison method. We have plotted a vertical line at the x-axis of 1, which is when there is an equal number of votes to comparisons. This location is a useful indicator, as it is generally speaking when an equal amount of resources are given to the two approaches, but should not be taken as the only fair comparison point. Instead, we understand it to be the comparison method's worst relative performance, as the majority-vote method tends to be more limited in terms of scaling with additional resources, which we show in Fig.~\ref{fig:observer-bias} (right panel). On the other hand, the number of possible comparisons exceeds the 3-vote majority-vote for $N>8$, moving us to the region of the figures to the right of $x=1$.

\subsubsection{Effects of Subjectivity}
In this section, we analyse the relative performance of the majority-vote method and the comparison method by measuring the difference in binary labelling performance, as indicated by $f_1$ scores. A positive value of difference in $f_1$ score indicates better performance of the comparison method in binary classification compared to the majority-vote method. To this end, we simulate 3-vote majority-voting and comparison tasks while varying values for rater threshold diversity, ambiguity in comparison, and ambiguity in perception. Sections~\ref{res:rater-bias} and \ref{res:additional-results} will extend these simulations to $2n+1$-vote majority-voting and rater populations with implicit biases.  

\begin{figure}[ht]
    \centering
    \includegraphics[width=0.5\linewidth]{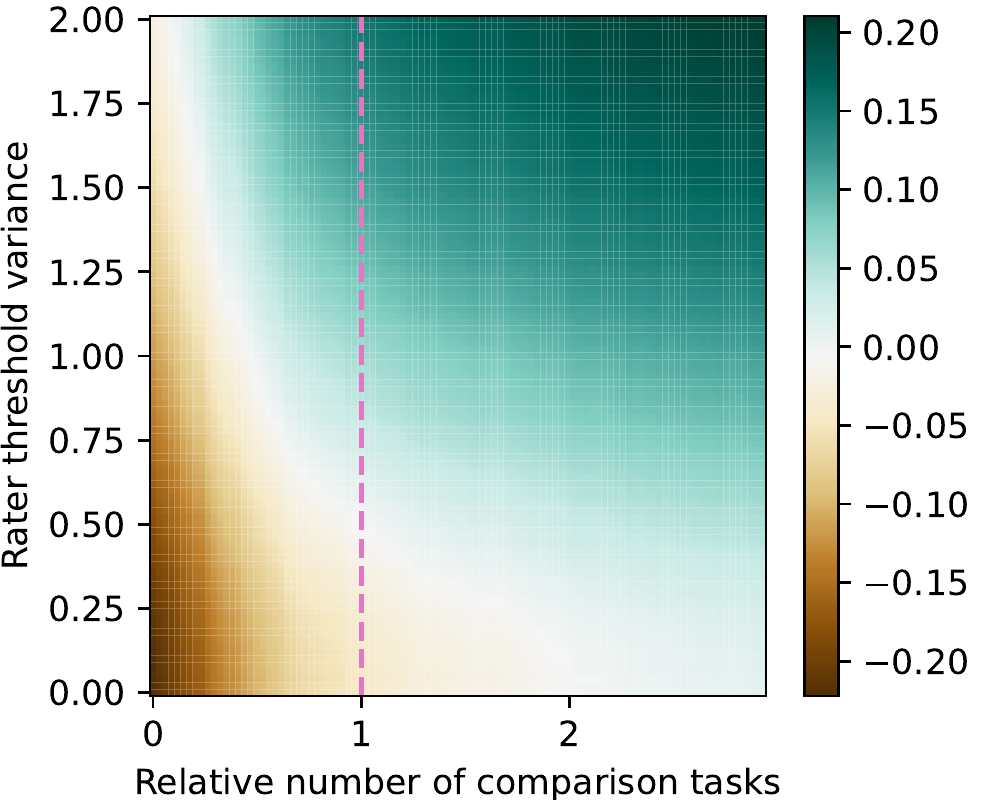}
    \caption{Subjectivity of a problem, expressed in terms of diversity of threshold among raters, has a positive effect on the relative efficacy of the pairwise comparison method. The heat map shows the difference in positive label $f_1$ scores of the pairwise comparison method and the majority-vote method. The blue regions on the heat map (i.e. positive-valued) display areas where the comparison method performs better than the majority-vote method. The dashed vertical line indicates the situation where equal numbers of comparison tasks and majority-vote tasks are performed.}
    \label{fig:just-threshold-diversity}
\end{figure}

\begin{figure}
    \centering
    \includegraphics[width=0.5\linewidth]{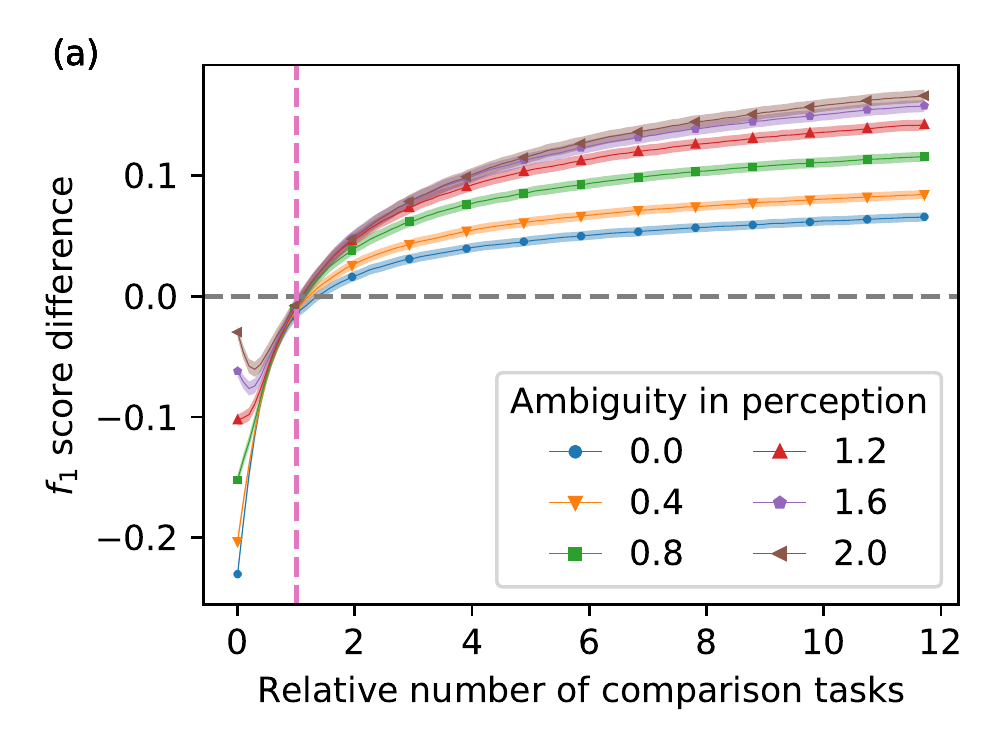}%
    \includegraphics[width=0.5\linewidth]{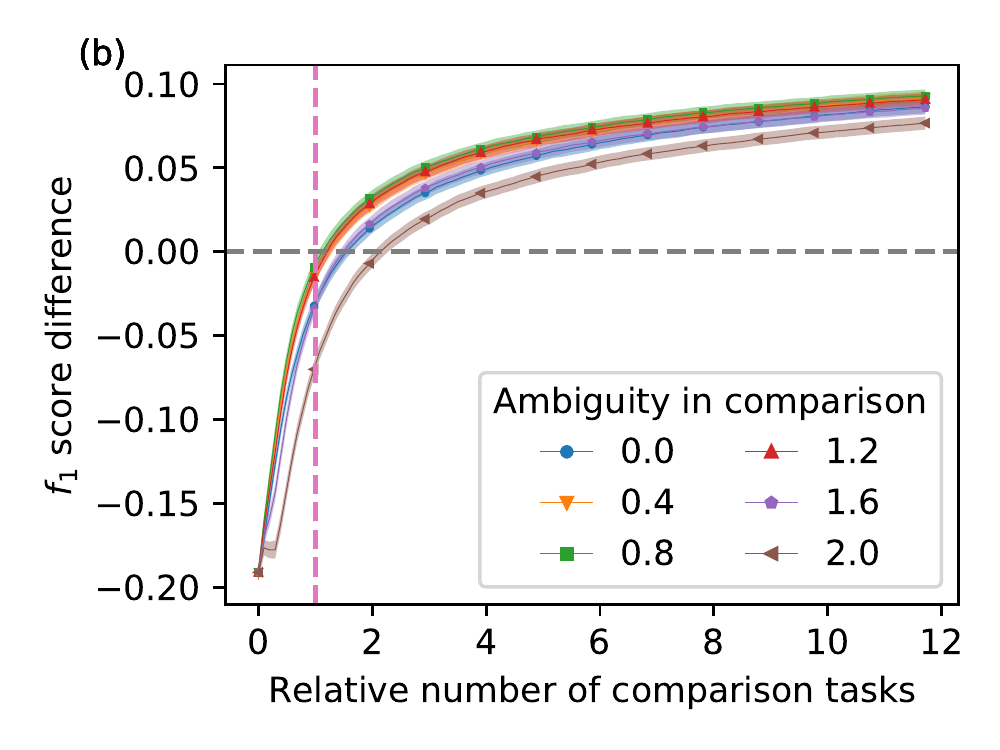}
    \includegraphics[width=0.5\linewidth]{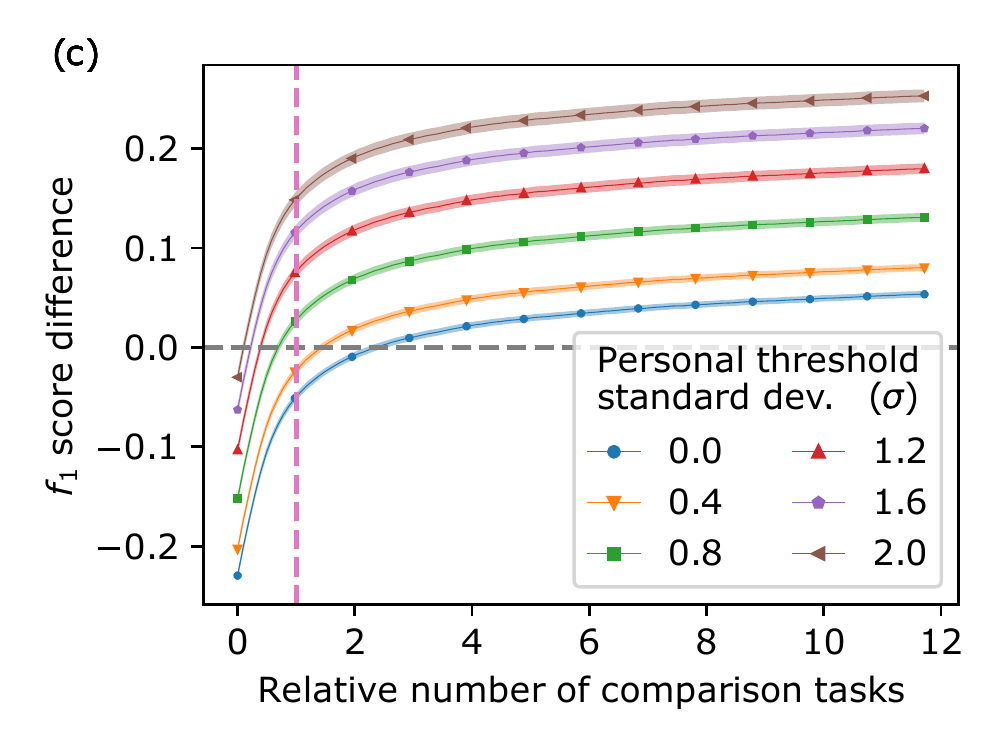}%
    \includegraphics[width=0.5\linewidth]{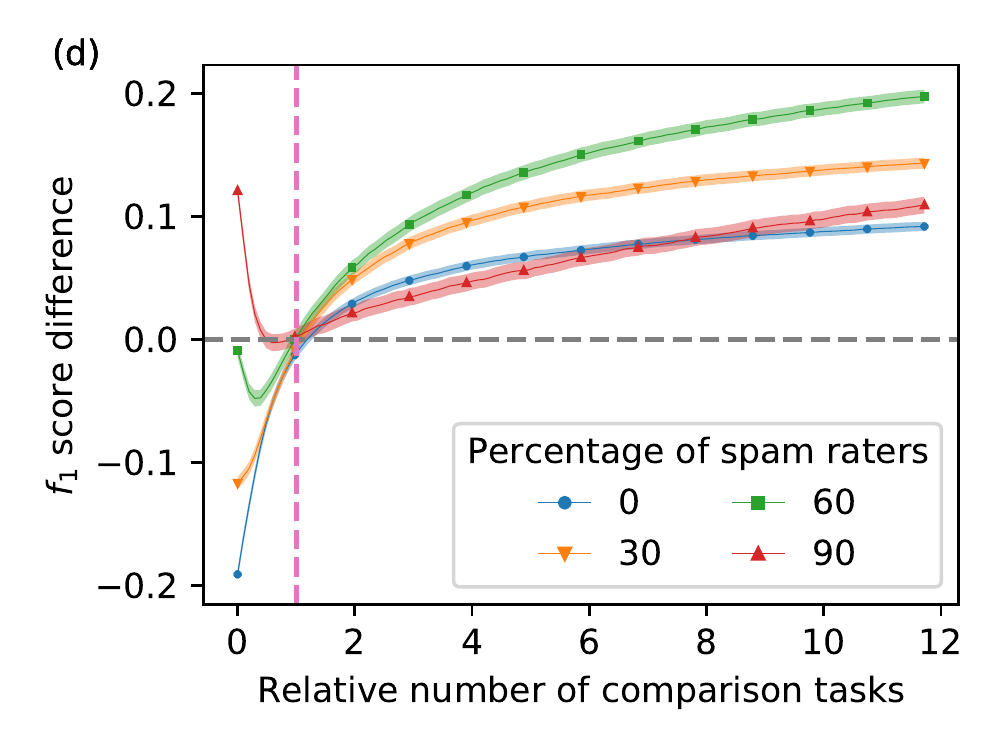}
    \caption{$f_1$ score difference ($f_1^\text{comparison} - f_1^\text{majority}$) of the comparison and majority-vote methods for parameters ambiguity in perception (a), ambiguity in comparison (b), personal threshold differences between raters (c) and percentage of spam raters (d) as the relative number of comparison tasks, that is, the number of comparison tasks divided by the number of the majority-vote tasks, increases. A positive $f_1$ score difference (shown on the y-axis) indicates an advantage in using the comparison method over the majority-vote method in terms of binary labelling accuracy with a given ratio of comparison tasks to majority-vote tasks. The x-axis indicates the ratio of comparison tasks to majority-vote tasks, i.e.~a value of 2 indicates performing twice as many comparisons as votes, where the number of majority-vote tasks is fixed to 3 tasks (votes) per item. The vertical line at x=1 indicates performing the same number of comparison tasks and majority-vote tasks. Larger values of ambiguity in perception and higher variance of personal threshold, as with medium amounts of ambiguity in comparison and spammer raters, result in a higher relative efficacy of the comparison method compared to the majority-voting method.}
    \label{fig:subjective_params}
\end{figure}

\paragraph{Diversity in threshold.}
We begin by looking at the first type of translational subjectivity, which is the diversity in raters' thresholds for identifying an item to be possessing the construct in question (e.g.~how much toxicity is toxic?) which is parameterised by the level of variability $\sigma$ in the threshold values in the rater population. As previously discussed, this type of translational subjectivity factors into item-wise rating tasks, and should therefore favour the comparison method over the majority-vote method. Our simulation results, presented in Fig.~\ref{fig:just-threshold-diversity}, confirm that higher degrees of subjectivity in the form of greater rater threshold diversity confers an advantage to the comparison method. The heat map shows the difference in the two methods' $f_1$ scores, with the blue regions indicating better performance by the comparison method. These findings are also presented in Fig.~\ref{fig:subjective_params}(c), which shows $f_1$ score differences at specific values of $\sigma$.

\paragraph{Ambiguity in comparison.}
Next, we consider the other type of translational subjectivity, ambiguity in comparison, which factors only into comparison tasks. It stands to reason that it should negatively impact the performance of the comparison method, therefore favouring the majority-vote method. This is indeed generally the case, as shown in Fig.~\ref{fig:subjective_params}(b). However, interestingly, we also see that a moderate amount of ambiguity in comparison actually improves the accuracy of the comparison method. Specifically, the performance of the comparison method improves as the ambiguity in comparison is increased to 0.8, after which its performance drops back down. This is due to the fact that in our implementation of the Elo rating system, when the raters are unable to differentiate between two items, they are allowed to declare them as equal, which results in the ratings of the two items converging toward each other. Past a certain point, however, large values of ambiguity in comparison mean even items with a large disparity in true ratings cannot be distinguished from each other, thereby resulting in less information made available to the rating system. For example, when the ambiguity in comparison has a value of 2, an item in the third percentile in terms of true rating is deemed indistinguishable from an item in the 97th percentile.

\paragraph{Ambiguity in perception.}\label{res:ambiguity-in-perception}
Turning to definitional subjectivity, Fig.~\ref{fig:subjective_params}(a) illustrates how the comparison and the majority-vote methods perform relative to each other for six different values of ambiguity in perception. We see that when there is an equal number of comparisons to votes, the two approaches perform the same. However, as discussed in Sec.~\ref{res:subjectivity}, the majority-vote method is capped at $3N$ tasks unless a larger number of votes are cast per item, which does not scale well. The region for our assessment, then, is to the right of the vertical pink line, where larger values of perception ambiguity result in the comparison method having a larger lead in accuracy to that of the majority-vote method. This lead is accentuated as the number of comparison tasks performed increases. 

\subsubsection{Rater bias}\label{res:rater-bias}
In this setup, each item has an additional binary observable representing a \textit{discriminatory feature} (e.g., racial features in the profile photo) which ideally should not be affecting the assessment of an item. The raters may, nonetheless, take it into consideration in their assessment. If the weight distribution associated with the binary feature has a mean of zero in the rater population, this will simply induce a larger error in estimated ratings for the items exhibiting that feature. On the other hand, bias is introduced to the classification if the raters are more likely to interpret the existence of this additional feature in one direction. This happens when the weight distribution associated with the discriminatory feature has a nonzero mean, e.g., on average seeing a black profile picture makes raters more likely to assess the comment as toxic regardless of the content.

\begin{figure}[ht]
    \centering
    \includegraphics[width=0.5\linewidth]{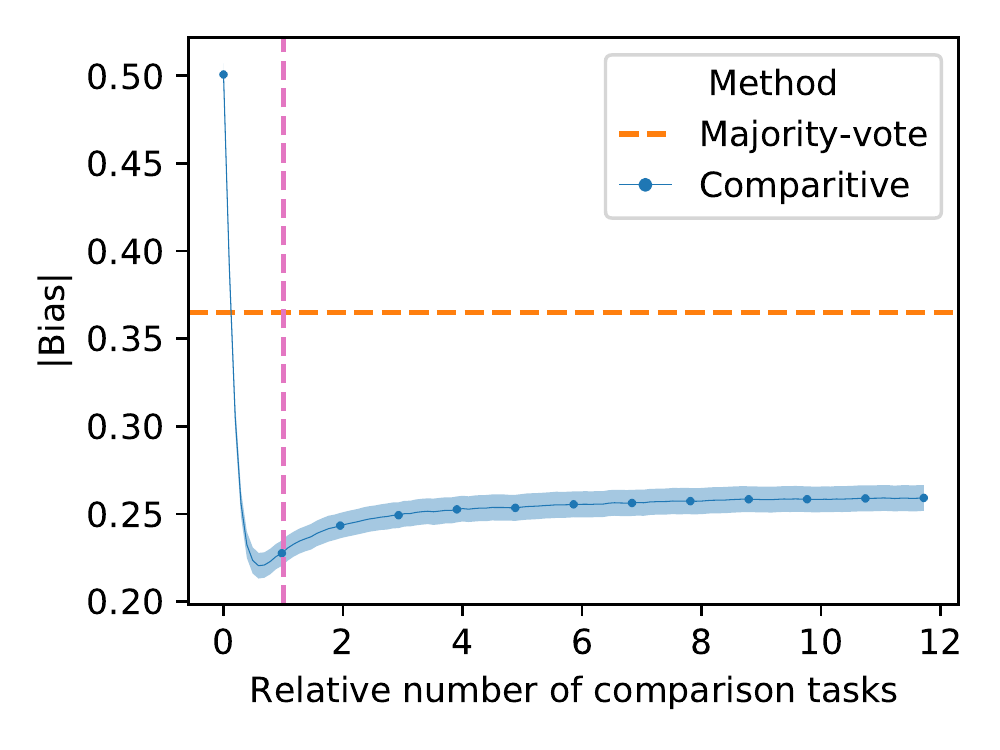}%
    \includegraphics[width=0.5\linewidth]{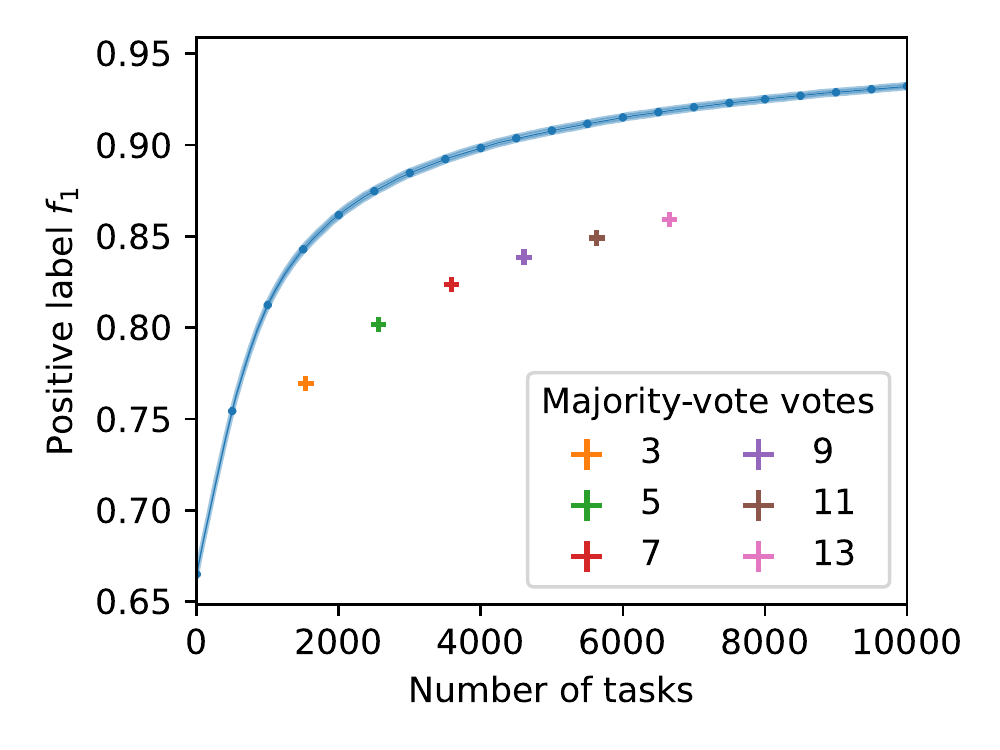}
    \caption{Left: The magnitude of bias in the binary classification of items with an observable discriminatory feature for the comparison and majority-vote methods with the same set of raters. The bias is calculated as the mean difference between the predicted label (using either the comparison or majority-vote method) and the true labels of all items that exhibit the biased feature. With the same group of raters, items exhibiting the discriminated feature are less likely to be classified with the negative label using the comparison method as compared to the majority-vote method. Right: The efficacy of the comparison method for a subjective task in a diverse population of simulated raters with personal threshold variance of 1.0, as compared to majority-voting method with different number of votes per item. The comparison method consistently out-performs the majority-vote method. As the number of votes of majority voting increases, the difference becomes slightly less substantial compared to the commonplace 3-vote majority voting.}
    \label{fig:observer-bias}
\end{figure}

To simulate this, we add to each item an additional binary ${0, 1}$ feature with probability $0.5$. The weight associated with this feature is distributed in population as a skewed beta distribution with parameters $\alpha=2$ and $\beta=16$ translated and scaled to have support of $[-1, +1]$, i.e.~ $\omega_d \sim 2 Beta(2, 16) - 1$. This implies that the raters are on average more likely to associate the exhibition of the discriminatory feature to lower ratings and therefore the negative label. The results in Fig.~\ref{fig:observer-bias} (left panel) show that while the comparison method still reflects the bias against the discriminatory feature that exists among the raters, the magnitude of the exhibited bias is generally quite smaller than in the majority-vote method.

\subsubsection{Additional results}\label{res:additional-results}
In addition to assessing the impact of subjectivity on the two methods, we conducted additional analysis on how the methods perform when faced with spam, and when the number of votes increases in the majority-vote method.

\paragraph{The effect of spam}
For this analysis, we simulated the effect of spammers by having a certain percentage of raters give random responses to the comparisons or absolute vote tasks. Our results are presented in Fig.~\ref{fig:subjective_params}(d). Our findings here are similar to the results for ambiguity in perception, where while the two methods perform the same when there is resource parity, the comparison method can scale, making the relevant region for our assessment to the right of the vertical pink line. Here, we see that the comparison method is more robust to spamming than the majority-vote. 

\paragraph{Increasing voters per majority-vote task}
It is possible to run the majority-vote method with more than 3 votes. Fig.~\ref{fig:observer-bias} (right panel) compares the expected $f_1$ scores of the majority-vote method with different numbers of votes per item, to those of the comparison method with the same number of tasks. 

Through the simulations analysed in this section, we are able to see how the comparison method outperforms the majority-vote in the increasing presence of subjectivity, namely with diversity in threshold and ambiguity in perception. Further, we find that while higher ambiguity in comparison has an adverse effect on the performance of the comparison method, a moderate amount in fact proves to be beneficial. We subsequently showed that the comparison method is more robust to rater bias and spam when compared to the majority-vote method. 

There is nevertheless a trade-off between the accuracy of the comparison method and the number of pairwise comparisons. We study this trade-off and how the number of required comparisons and labelling accuracy scale with the total number of items in the following section.

\subsection{Scaling behaviour}\label{res:scaling}
To understand the scaling behaviour of the comparison method, we can compare the labelling accuracy of the comparison method both in simulation and in empirical observation with different numbers of items. Take for example a set of $N$ items. For different numbers of comparisons ($n_\text{comparisons}$) we plot the average positive label $f_1$ score. If we repeat the same for different values of $N$, we arrive at a family of trajectories. These trajectories collapse on top of each other after applying a scaling function of $N$ to the number of comparisons (the horizontal axis) for each curve. This scaling function can tell us how one needs to increase the number of comparisons when increasing the system size $N$ for achieving the same level of quality.

\begin{figure}
    \centering
    \includegraphics[width=\linewidth]{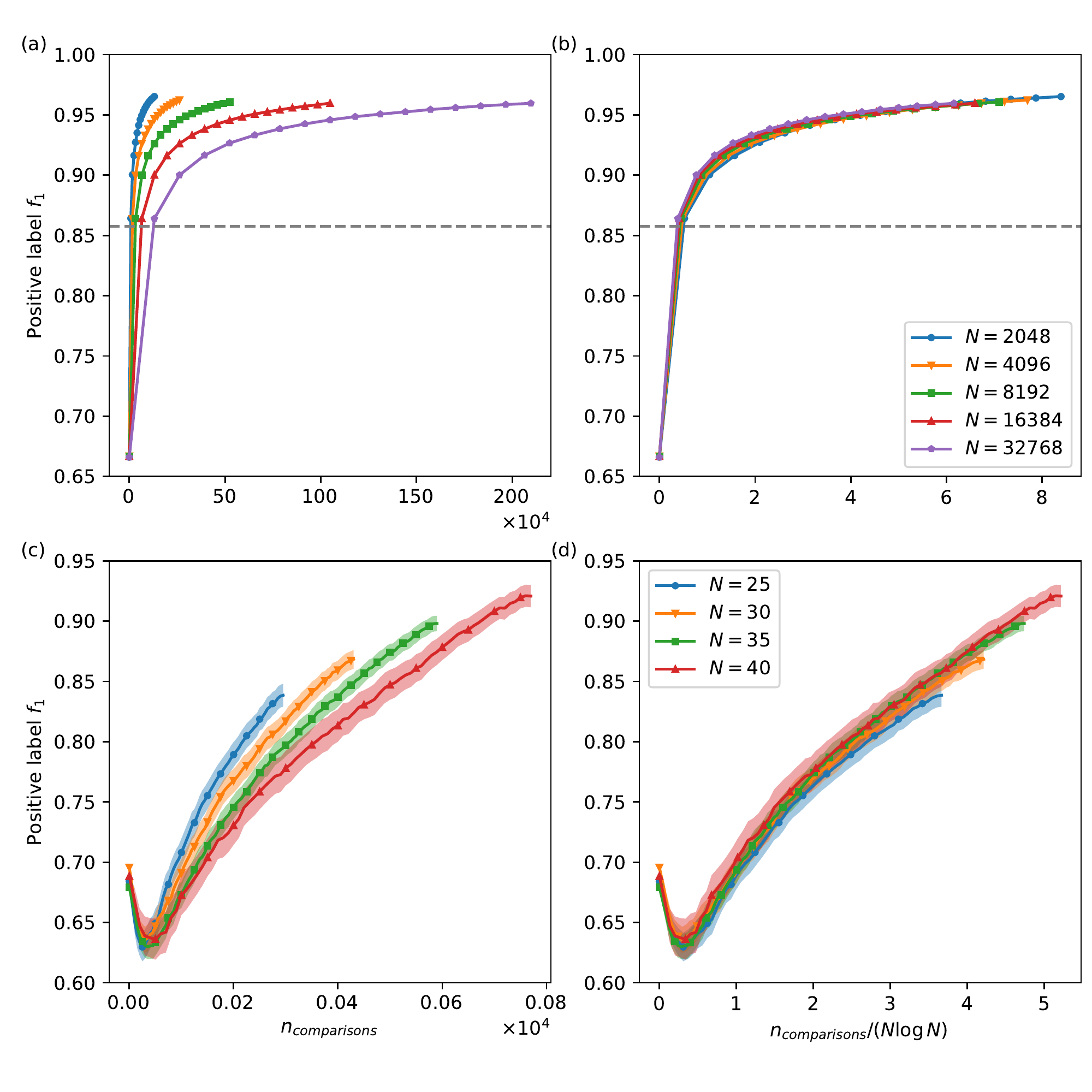}
    \caption{The number of required comparisons for achieving desired level of accuracy, as denoted by the positive label $f_1$ score, in a subjective binary classification task scales log-linearly $O(N \log N)$ with the number of items being classified with the comparison method $N$. This can be verified by the scaling analysis on (b) simulations described in Sec.~\ref{sec:simulation} and (d) empirical analysis on real-world subjective task described in Sec.~\ref{sec:experiments}. The collapse of the trajectories, after correcting for the hypothesised scaling function $N \log N$ on the horizontal axis, validates the hypothesised scaling function. The hyphenated horizontal lines in (a) and (b) denote the expected positive label $f_1$ score for the 3-vote majority-voting classification method. The shaded area indicates the 5-$\sigma$ confidence interval where $\sigma$ is the standard error of mean.}
    \label{fig:scaling}
\end{figure}

The same analysis was repeated on the experimental data described in Sec.~\ref{sec:experiments}, with the key difference being that true ratings and labels are not available for the experimental items. Instead, we used the Elo rating based on the full set of all 1640 performed comparisons as the comparison baseline. Different system sizes can be simulated by selecting a subset of items with size $N$, and removing all comparisons that include items not in this subset. For each system size, positive and negative ``true'' labels were assigned based on comparing the full-set rating of each item in the subset with the subset median rating. For each system size, many random samples of comparisons with different size $n_\text{comparisons}$ were selected, and classification labels were predicted based on Elo ratings calculated from these comparisons. For each system size, this process is repeated many times. Figure \ref{fig:scaling}(c) shows the change in $f_1$ score as a function of $n_\text{comparisons}$ and $N$, while Fig.~\ref{fig:scaling}(d) shows that the same trajectories collapse after correcting for our hypothesised $O(N \log N)$ scaling, confirming the hypothesis that the number of required comparisons for the same level of labelling quality grows log-linearly with respect to the number of items.

These results, as well as those based on the mathematical modelling presented in Sec.~\ref{res:subjectivity}, lay a road-map for using our proposed comparison method for crowdsourcing tasks intended to measure subjective constructs. In the next section, we summarise our results and present our recommendations for real-world implementation of our proposed method.

\section{Discussion and conclusion}
In this paper, we proposed a comparison-based labelling approach for subjective constructs which combines pairwise comparisons with Elo scoring. We showed through simulations that the comparison method fares better against the ubiquitous majority-vote method in dealing with most types of subjectivity, measured as reductions in misclassification error and classification bias.

Specifically, using an agent-based model of crowdsourced labelling to quantify different forms of task subjectivity, we compared the performance of the two labelling approaches. We found that the comparison method either is better at reducing random error (i.e. achieving higher $f_1$ scores) when the number of comparisons and votes are equal (i.e. under resource parity), or has more room to improve with an increasing number of tasks whereas the majority-vote tends to become capped. Second, we incorporated a discriminatory feature into our agent-based model to test how the two approaches performed in bias reduction. We find a nonzero value of bias in both methods, which demonstrates that bias reduction needs to be accomplished via other steps in the human-labelling pipeline. However, the comparison method outperforms the majority-vote in that it does not inflate bias in the way majority-voting does.

\subsection{Considerations when using the comparison method in real-world studies}\label{sec:application}
To facilitate applied work, we conducted a set of comparison-based crowdsourcing tasks with Twitter conversations to study how the number of comparison-based tasks scales with respect to the number of to-be-labelled items. Based on this experiment, we arrived at an $O(N \log N)$ scaling of the required random comparisons. Our results show that the previously assumed relationship between the number of comparison tasks and items in the literature (i.e. $O(N^2)$) overestimated the resource intensity of the comparison method \cite{bagrow2020democratizing}. Instead, our observed log-linear scaling means that comparison-based labelling is more tractable in large-scale studies than previously thought.\footnote{This allows researchers to run small-scale pilots until the desired accuracy of results are reached, after which comparison tasks can be scaled up to an arbitrary number of items, with the number of required comparisons growing as $O(N \log N)$ for the same accuracy level.}

In Appendix~\ref{appendix:application}, we provide a step-by-step guide for implementing our proposed pipeline and also introduce our open-source reference implementation of the Elo rating system. Here, we briefly discuss some additional considerations that concern crowdsourced labelling tasks in general. 

First, regardless of which labelling method, it is important to keep in mind the representativeness of the crowd worker sample to the desired population. Following our discussion in Sec.~\ref{sec:theory}, the target measure when working with subjective constructs is the mean population perception $\mu_\theta$ of a given item, which is sensitive to sampling bias. We did not address the question of sampling bias in our study, but point to the survey sampling literature that provides instruction on how to address these concerns \citep{gelman1997poststratification,gao2021improving}.

Second, subjectivity-induced variation aside, crowdsourced labels can vary in quality so quality control is an important consideration. In our study we explored the effect of non-malicious spam and otherwise random labels, which enters into our theoretical framework in the same manner as ambiguity in perception. We show that spam mitigation for the comparison method scales better with resources relative to the majority-vote. We also discuss spam detection methods in Appendix~\ref{appendix:spam}. We do not address how to account for crowd raters who tactically attempt to sway the results, which we believe should be minimal for innocuous tasks where there is no internal incentive for maliciousness. However, we recognise it is a potentially important consideration when working with politically sensitive topics. 

Related to this final point, we note that our proposed pipeline is modular, which makes it easy for components to be customised for the specific needs of a study. For example, subsystems designed to detect and mitigate intentionally malicious workers can be plugged into the pipeline between the labelling and aggregation tasks in a straightforward manner.

\subsection{Limitations and future work}
We conclude by discussing avenues of future work that address limitations of our current study. One main area of future exploration stem from our synthetic design and largely modular testing framework. First, we have tested different types of subjectivity independently of each other instead of also considering how they may interact. Measuring and testing potential interactions between various aspects of subjectivity we delineate in our model is a difficult task and will requiring novel study design and a broad data-gathering effort. Part of this requires future work to move toward real-world data, which lets us assess the holistic performance of different crowdsource labelling methods, as it is difficult to directly compare the magnitude of different types of subjectivity in a synthetic setting.

Relatedly, by focusing on only the labelling and aggregation components, we have largely assumed away how other important considerations in a crowdsource labelling pipeline (e.g. crowdworker representativeness, bias reduction, and spam and gaming mitigation) can interact with the labelling and aggregation steps. For example, in terms of bias reduction, although we show that the comparison method is less susceptible to discriminatory features compared to the majority-vote method, we do not propose a solution to correct it. An active measure might be required on part of the researcher to eliminate or account for the biases introduced by the possibly discriminatory features of the items. This is an active area of research \citep{wallace2022debiased,li2020towards}, which we believe can be fruitfully combined with work such as ours that focus on labelling and aggregation in crowdsourcing tasks.

Some additional extensions of our work are worth exploring. First, while we tried to include many aspects of subjectivity in our simple mathematical model of human understanding, there may exist other facets to subjectivity that we have not incorporated into our formalisation. For example, in this study, we assumed unimodal distributions for weights, which recent work has expanded beyond \citep{aroyo2015truth}. Second, we do not provide analytical proof of the scaling relationship $O(N \log N)$. Rather, we took an empirical approach to arrive at a scaling relationship. We believe that an analytical derivation of this scaling relationship would prove a valuable addition to this discussion. Finally, while we worked with binary measures in this study, the comparison method and the Elo rating system directly allow for ordinal or even interval measures. These can be used instead of the binary label to train a regression model or to study the correlation of item ratings with time or across groups. As such, our work likely understates the performance of the comparison method relative to item-wise labelling which requires extended task designs to allow for nonbinary labels, and are likely prone to higher levels of translational subjectivity. Much more work can be done in this area to understand these potential outputs and how they perform in capturing the target subjective construct. We hope our present work provides a useful road-map for future work.

\begin{acks}
We thank Hande Celikkanat, Tuomo Hiippala, Mikko Kivel\"a, Kevin Reuning, Rosa Suviranta, and Tuomas Yl\"a-Anttila for their helpful input at various points in this project. This project is supported by Academy of Finland grants 320780 and 320781, and the 2021 Helsinki Institute of Social Science and Humanities Catalyst Grant Funding. We acknowledge the computational resources provided by the Aalto Science-IT project.
\end{acks}

\bibliographystyle{ACM-Reference-Format}
\bibliography{comparison_method_references}

%%% -*-BibTeX-*-
%%% Do NOT edit. File created by BibTeX with style
%%% ACM-Reference-Format-Journals [18-Jan-2012].

\begin{thebibliography}{65}

%%% ====================================================================
%%% NOTE TO THE USER: you can override these defaults by providing
%%% customized versions of any of these macros before the \bibliography
%%% command.  Each of them MUST provide its own final punctuation,
%%% except for \shownote{}, \showDOI{}, and \showURL{}.  The latter two
%%% do not use final punctuation, in order to avoid confusing it with
%%% the Web address.
%%%
%%% To suppress output of a particular field, define its macro to expand
%%% to an empty string, or better, \unskip, like this:
%%%
%%% \newcommand{\showDOI}[1]{\unskip}   % LaTeX syntax
%%%
%%% \def \showDOI #1{\unskip}           % plain TeX syntax
%%%
%%% ====================================================================

\ifx \showCODEN    \undefined \def \showCODEN     #1{\unskip}     \fi
\ifx \showDOI      \undefined \def \showDOI       #1{#1}\fi
\ifx \showISBNx    \undefined \def \showISBNx     #1{\unskip}     \fi
\ifx \showISBNxiii \undefined \def \showISBNxiii  #1{\unskip}     \fi
\ifx \showISSN     \undefined \def \showISSN      #1{\unskip}     \fi
\ifx \showLCCN     \undefined \def \showLCCN      #1{\unskip}     \fi
\ifx \shownote     \undefined \def \shownote      #1{#1}          \fi
\ifx \showarticletitle \undefined \def \showarticletitle #1{#1}   \fi
\ifx \showURL      \undefined \def \showURL       {\relax}        \fi
% The following commands are used for tagged output and should be
% invisible to TeX
\providecommand\bibfield[2]{#2}
\providecommand\bibinfo[2]{#2}
\providecommand\natexlab[1]{#1}
\providecommand\showeprint[2][]{arXiv:#2}

\bibitem[Alonso et~al\mbox{.}(2015)]%
        {alonso2015debugging}
\bibfield{author}{\bibinfo{person}{Omar Alonso}, \bibinfo{person}{Catherine~C
  Marshall}, {and} \bibinfo{person}{Marc Najork}.}
  \bibinfo{year}{2015}\natexlab{}.
\newblock \showarticletitle{Debugging a crowdsourced task with low inter-rater
  agreement}. In \bibinfo{booktitle}{\emph{Proceedings of the 15th ACM/IEEE-CS
  Joint Conference on Digital Libraries}}. \bibinfo{pages}{101--110}.
\newblock
\urldef\tempurl%
\url{https://doi.org/10.1145/2756406.2757741}
\showDOI{\tempurl}


\bibitem[Aroyo et~al\mbox{.}(2019)]%
        {aroyo2019crowdsourcing}
\bibfield{author}{\bibinfo{person}{Lora Aroyo}, \bibinfo{person}{Lucas Dixon},
  \bibinfo{person}{Nithum Thain}, \bibinfo{person}{Olivia Redfield}, {and}
  \bibinfo{person}{Rachel Rosen}.} \bibinfo{year}{2019}\natexlab{}.
\newblock \showarticletitle{Crowdsourcing Subjective Tasks: The Case Study of
  Understanding Toxicity in Online Discussions}. In
  \bibinfo{booktitle}{\emph{Companion of The 2019 World Wide Web Conference,
  {WWW} 2019, San Francisco, CA, USA, May 13-17, 2019}}.
  \bibinfo{publisher}{{ACM}}, \bibinfo{pages}{1100--1105}.
\newblock
\urldef\tempurl%
\url{https://doi.org/10.1145/3308560.3317083}
\showDOI{\tempurl}


\bibitem[Aroyo and Welty(2015)]%
        {aroyo2015truth}
\bibfield{author}{\bibinfo{person}{Lora Aroyo} {and} \bibinfo{person}{Chris
  Welty}.} \bibinfo{year}{2015}\natexlab{}.
\newblock \showarticletitle{Truth is a lie: Crowd truth and the seven myths of
  human annotation}.
\newblock \bibinfo{journal}{\emph{AI Magazine}} \bibinfo{volume}{36},
  \bibinfo{number}{1} (\bibinfo{year}{2015}), \bibinfo{pages}{15--24}.
\newblock
\urldef\tempurl%
\url{https://doi.org/10.1609/aimag.v36i1.2564}
\showDOI{\tempurl}


\bibitem[Bagrow(2020)]%
        {bagrow2020democratizing}
\bibfield{author}{\bibinfo{person}{James~P Bagrow}.}
  \bibinfo{year}{2020}\natexlab{}.
\newblock \showarticletitle{Democratizing AI: non-expert design of prediction
  tasks}.
\newblock \bibinfo{journal}{\emph{PeerJ Computer Science}}  \bibinfo{volume}{6}
  (\bibinfo{year}{2020}), \bibinfo{pages}{e296}.
\newblock
\urldef\tempurl%
\url{https://doi.org/10.7717/peerj-cs.296}
\showDOI{\tempurl}


\bibitem[Balayn and Bozzon(2019)]%
        {balayn2019designing}
\bibfield{author}{\bibinfo{person}{Agathe Balayn} {and}
  \bibinfo{person}{Alessandro Bozzon}.} \bibinfo{year}{2019}\natexlab{}.
\newblock \showarticletitle{Designing evaluations of machine learning models
  for subjective inference: the case of sentence toxicity}.
\newblock \bibinfo{journal}{\emph{arXiv preprint}} (\bibinfo{year}{2019}).
\newblock
\urldef\tempurl%
\url{https://doi.org/10.48550/arXiv.1911.02471}
\showDOI{\tempurl}


\bibitem[Borji(2019)]%
        {borji2019pros}
\bibfield{author}{\bibinfo{person}{Ali Borji}.}
  \bibinfo{year}{2019}\natexlab{}.
\newblock \showarticletitle{Pros and cons of gan evaluation measures}.
\newblock \bibinfo{journal}{\emph{Computer Vision and Image Understanding}}
  \bibinfo{volume}{179} (\bibinfo{year}{2019}), \bibinfo{pages}{41--65}.
\newblock


\bibitem[Cabrera et~al\mbox{.}(2014)]%
        {cabrera2014systematic}
\bibfield{author}{\bibinfo{person}{Guillermo~F Cabrera},
  \bibinfo{person}{Christopher~J Miller}, {and} \bibinfo{person}{Jeff
  Schneider}.} \bibinfo{year}{2014}\natexlab{}.
\newblock \showarticletitle{Systematic labeling bias: De-biasing where everyone
  is wrong}. In \bibinfo{booktitle}{\emph{2014 22nd International Conference on
  Pattern Recognition}}. IEEE, \bibinfo{pages}{4417--4422}.
\newblock


\bibitem[Carterette et~al\mbox{.}(2008)]%
        {carterette2008here}
\bibfield{author}{\bibinfo{person}{Ben Carterette}, \bibinfo{person}{Paul~N.
  Bennett}, \bibinfo{person}{David~Maxwell Chickering}, {and}
  \bibinfo{person}{Susan~T. Dumais}.} \bibinfo{year}{2008}\natexlab{}.
\newblock \showarticletitle{Here or There}. In
  \bibinfo{booktitle}{\emph{Advances in Information Retrieval, 30th European
  Conference on {IR} Research, {ECIR} 2008, Glasgow, UK, March 30-April 3,
  2008. Proceedings}} \emph{(\bibinfo{series}{Lecture Notes in Computer
  Science}, Vol.~\bibinfo{volume}{4956})}. \bibinfo{publisher}{Springer},
  \bibinfo{pages}{16--27}.
\newblock
\urldef\tempurl%
\url{https://doi.org/10.1007/978-3-540-78646-7\_5}
\showDOI{\tempurl}


\bibitem[Chang et~al\mbox{.}(2017)]%
        {chang2017revolt}
\bibfield{author}{\bibinfo{person}{Joseph~Chee Chang}, \bibinfo{person}{Saleema
  Amershi}, {and} \bibinfo{person}{Ece Kamar}.}
  \bibinfo{year}{2017}\natexlab{}.
\newblock \showarticletitle{Revolt: Collaborative crowdsourcing for labeling
  machine learning datasets}. In \bibinfo{booktitle}{\emph{Proceedings of the
  2017 CHI Conference on Human Factors in Computing Systems}}.
  \bibinfo{pages}{2334--2346}.
\newblock
\urldef\tempurl%
\url{https://doi.org/10.1145/3025453.3026044}
\showDOI{\tempurl}


\bibitem[Chen et~al\mbox{.}(2013)]%
        {chen2013pairwise}
\bibfield{author}{\bibinfo{person}{Xi Chen}, \bibinfo{person}{Paul~N Bennett},
  \bibinfo{person}{Kevyn Collins-Thompson}, {and} \bibinfo{person}{Eric
  Horvitz}.} \bibinfo{year}{2013}\natexlab{}.
\newblock \showarticletitle{Pairwise ranking aggregation in a crowdsourced
  setting}. In \bibinfo{booktitle}{\emph{Proceedings of the sixth ACM
  international conference on Web search and data mining}}.
  \bibinfo{pages}{193--202}.
\newblock
\urldef\tempurl%
\url{https://doi.org/10.1145/2433396.2433420}
\showDOI{\tempurl}


\bibitem[Cheng and Bernstein(2015)]%
        {cheng2015flock}
\bibfield{author}{\bibinfo{person}{Justin Cheng} {and}
  \bibinfo{person}{Michael~S Bernstein}.} \bibinfo{year}{2015}\natexlab{}.
\newblock \showarticletitle{Flock: Hybrid crowd-machine learning classifiers}.
  In \bibinfo{booktitle}{\emph{Proceedings of the 18th ACM conference on
  computer supported cooperative work \& social computing}}.
  \bibinfo{pages}{600--611}.
\newblock
\urldef\tempurl%
\url{https://doi.org/10.1145/2675133.2675214}
\showDOI{\tempurl}


\bibitem[Chung et~al\mbox{.}(2019)]%
        {chung2019efficient}
\bibfield{author}{\bibinfo{person}{John Joon~Young Chung},
  \bibinfo{person}{Jean~Y Song}, \bibinfo{person}{Sindhu Kutty},
  \bibinfo{person}{Sungsoo Hong}, \bibinfo{person}{Juho Kim}, {and}
  \bibinfo{person}{Walter~S Lasecki}.} \bibinfo{year}{2019}\natexlab{}.
\newblock \showarticletitle{Efficient elicitation approaches to estimate
  collective crowd answers}.
\newblock \bibinfo{journal}{\emph{Proceedings of the ACM on Human-Computer
  Interaction}} \bibinfo{volume}{3}, \bibinfo{number}{CSCW}
  (\bibinfo{year}{2019}), \bibinfo{pages}{1--25}.
\newblock
\urldef\tempurl%
\url{https://doi.org/10.1145/3359164}
\showDOI{\tempurl}


\bibitem[Denton et~al\mbox{.}(2015)]%
        {denton2015deep}
\bibfield{author}{\bibinfo{person}{Emily~L Denton}, \bibinfo{person}{Soumith
  Chintala}, \bibinfo{person}{Rob Fergus}, {et~al\mbox{.}}}
  \bibinfo{year}{2015}\natexlab{}.
\newblock \showarticletitle{Deep generative image models using a Laplacian
  pyramid of adversarial networks}.
\newblock \bibinfo{journal}{\emph{Advances in neural information processing
  systems}}  \bibinfo{volume}{28} (\bibinfo{year}{2015}),
  \bibinfo{pages}{1486--1494}.
\newblock


\bibitem[Difallah et~al\mbox{.}(2012)]%
        {difallah2012mechanical}
\bibfield{author}{\bibinfo{person}{Djellel~Eddine Difallah},
  \bibinfo{person}{Gianluca Demartini}, {and} \bibinfo{person}{Philippe
  Cudr{\'e}-Mauroux}.} \bibinfo{year}{2012}\natexlab{}.
\newblock \showarticletitle{Mechanical cheat: Spamming schemes and adversarial
  techniques on crowdsourcing platforms}. In
  \bibinfo{booktitle}{\emph{CrowdSearch}} \emph{(\bibinfo{series}{{CEUR}
  Workshop Proceedings}, Vol.~\bibinfo{volume}{842})}.
  \bibinfo{publisher}{CEUR-WS.org}, \bibinfo{pages}{26--30}.
\newblock


\bibitem[Dixon et~al\mbox{.}(2018)]%
        {dixon2018measuring}
\bibfield{author}{\bibinfo{person}{Lucas Dixon}, \bibinfo{person}{John Li},
  \bibinfo{person}{Jeffrey Sorensen}, \bibinfo{person}{Nithum Thain}, {and}
  \bibinfo{person}{Lucy Vasserman}.} \bibinfo{year}{2018}\natexlab{}.
\newblock \showarticletitle{Measuring and mitigating unintended bias in text
  classification}. In \bibinfo{booktitle}{\emph{Proceedings of the 2018
  AAAI/ACM Conference on AI, Ethics, and Society}}. \bibinfo{publisher}{{ACM}},
  \bibinfo{pages}{67--73}.
\newblock
\urldef\tempurl%
\url{https://doi.org/10.1145/3278721.3278729}
\showDOI{\tempurl}


\bibitem[Drapeau et~al\mbox{.}(2016)]%
        {drapeau2016microtalk}
\bibfield{author}{\bibinfo{person}{Ryan Drapeau}, \bibinfo{person}{Lydia
  Chilton}, \bibinfo{person}{Jonathan Bragg}, {and} \bibinfo{person}{Daniel
  Weld}.} \bibinfo{year}{2016}\natexlab{}.
\newblock \showarticletitle{Microtalk: Using argumentation to improve
  crowdsourcing accuracy}. In \bibinfo{booktitle}{\emph{Proceedings of the AAAI
  Conference on Human Computation and Crowdsourcing}},
  Vol.~\bibinfo{volume}{4}. \bibinfo{pages}{32--41}.
\newblock
\urldef\tempurl%
\url{https://doi.org/10.1609/hcomp.v4i1.13270}
\showDOI{\tempurl}


\bibitem[Dumitrache(2015)]%
        {dumitrache2015crowdsourcing}
\bibfield{author}{\bibinfo{person}{Anca Dumitrache}.}
  \bibinfo{year}{2015}\natexlab{}.
\newblock \showarticletitle{Crowdsourcing disagreement for collecting semantic
  annotation}. In \bibinfo{booktitle}{\emph{European Semantic Web Conference}}.
  Springer, \bibinfo{pages}{701--710}.
\newblock
\urldef\tempurl%
\url{https://doi.org/10.1007/978-3-319-18818-8_43}
\showDOI{\tempurl}


\bibitem[Dumitrache et~al\mbox{.}(2017)]%
        {dumitrache2017crowdsourcing}
\bibfield{author}{\bibinfo{person}{Anca Dumitrache}, \bibinfo{person}{Lora
  Aroyo}, {and} \bibinfo{person}{Chris Welty}.}
  \bibinfo{year}{2017}\natexlab{}.
\newblock \showarticletitle{Crowdsourcing ground truth for medical relation
  extraction}.
\newblock \bibinfo{journal}{\emph{arXiv preprint arXiv:1701.02185}}
  (\bibinfo{year}{2017}).
\newblock
\urldef\tempurl%
\url{https://doi.org/10.1145/3152889}
\showDOI{\tempurl}


\bibitem[Dumitrache et~al\mbox{.}(2018)]%
        {dumitrache2018capturing}
\bibfield{author}{\bibinfo{person}{Anca Dumitrache}, \bibinfo{person}{Lora
  Aroyo}, {and} \bibinfo{person}{Chris Welty}.}
  \bibinfo{year}{2018}\natexlab{}.
\newblock \showarticletitle{Capturing ambiguity in crowdsourcing frame
  disambiguation}. In \bibinfo{booktitle}{\emph{Sixth AAAI Conference on Human
  Computation and Crowdsourcing}}.
\newblock


\bibitem[Gao et~al\mbox{.}(2021)]%
        {gao2021improving}
\bibfield{author}{\bibinfo{person}{Yuxiang Gao}, \bibinfo{person}{Lauren
  Kennedy}, \bibinfo{person}{Daniel Simpson}, {and} \bibinfo{person}{Andrew
  Gelman}.} \bibinfo{year}{2021}\natexlab{}.
\newblock \showarticletitle{Improving multilevel regression and
  poststratification with structured priors}.
\newblock \bibinfo{journal}{\emph{Bayesian Analysis}} \bibinfo{volume}{16},
  \bibinfo{number}{3} (\bibinfo{year}{2021}), \bibinfo{pages}{719--744}.
\newblock


\bibitem[Gelman and Little(1997)]%
        {gelman1997poststratification}
\bibfield{author}{\bibinfo{person}{Andrew Gelman} {and}
  \bibinfo{person}{Thomas~C Little}.} \bibinfo{year}{1997}\natexlab{}.
\newblock \showarticletitle{Poststratification into many categories using
  hierarchical logistic regression}.
\newblock  (\bibinfo{year}{1997}).
\newblock


\bibitem[Gentner and Markman(1997)]%
        {gentner1997structure}
\bibfield{author}{\bibinfo{person}{Dedre Gentner} {and}
  \bibinfo{person}{Arthur~B Markman}.} \bibinfo{year}{1997}\natexlab{}.
\newblock \showarticletitle{Structure mapping in analogy and similarity}.
\newblock \bibinfo{journal}{\emph{American Psychologist}} \bibinfo{volume}{52},
  \bibinfo{number}{1} (\bibinfo{year}{1997}), \bibinfo{pages}{45}.
\newblock
\urldef\tempurl%
\url{https://doi.org/10.1037/0003-066X.52.1.45}
\showDOI{\tempurl}


\bibitem[Goffin and Olson(2011)]%
        {goffin2011all}
\bibfield{author}{\bibinfo{person}{Richard~D Goffin} {and}
  \bibinfo{person}{James~M Olson}.} \bibinfo{year}{2011}\natexlab{}.
\newblock \showarticletitle{Is it all relative? Comparative judgments and the
  possible improvement of self-ratings and ratings of others}.
\newblock \bibinfo{journal}{\emph{Perspectives on Psychological Science}}
  \bibinfo{volume}{6}, \bibinfo{number}{1} (\bibinfo{year}{2011}),
  \bibinfo{pages}{48--60}.
\newblock


\bibitem[Goldberg(2010)]%
        {goldberg2010two}
\bibfield{author}{\bibinfo{person}{William~T Goldberg}.}
  \bibinfo{year}{2010}\natexlab{}.
\newblock \showarticletitle{Two nations, one web: Comparative legal approaches
  to pornographic obscenity by the United States and the United Kingdom}.
\newblock \bibinfo{journal}{\emph{BUL Rev.}}  \bibinfo{volume}{90}
  (\bibinfo{year}{2010}), \bibinfo{pages}{2121}.
\newblock


\bibitem[Guo et~al\mbox{.}(2018)]%
        {guo2018experimental}
\bibfield{author}{\bibinfo{person}{Yuan Guo}, \bibinfo{person}{Peng Tian},
  \bibinfo{person}{Jayashree Kalpathy{-}Cramer}, \bibinfo{person}{Susan Ostmo},
  \bibinfo{person}{J.~Peter Campbell}, \bibinfo{person}{Michael~F. Chiang},
  \bibinfo{person}{Deniz Erdogmus}, \bibinfo{person}{Jennifer~G. Dy}, {and}
  \bibinfo{person}{Stratis Ioannidis}.} \bibinfo{year}{2018}\natexlab{}.
\newblock \showarticletitle{Experimental Design under the Bradley-Terry Model}.
  In \bibinfo{booktitle}{\emph{Proceedings of the Twenty-Seventh International
  Joint Conference on Artificial Intelligence, {IJCAI} 2018, July 13-19, 2018,
  Stockholm, Sweden}}. \bibinfo{publisher}{ijcai.org},
  \bibinfo{pages}{2198--2204}.
\newblock
\urldef\tempurl%
\url{https://doi.org/10.24963/ijcai.2018/304}
\showDOI{\tempurl}


\bibitem[Hettiachchi et~al\mbox{.}(2020)]%
        {hettiachchi2020augmenting}
\bibfield{author}{\bibinfo{person}{Danula Hettiachchi}, \bibinfo{person}{Niels
  van Berkel}, \bibinfo{person}{Simo Hosio}, \bibinfo{person}{Miguel~Bordallo
  L{\'o}pez}, \bibinfo{person}{Vassilis Kostakos}, {and} \bibinfo{person}{Jorge
  Goncalves}.} \bibinfo{year}{2020}\natexlab{}.
\newblock \showarticletitle{Augmenting Automated Kinship Verification with
  Targeted Human Input}. In \bibinfo{booktitle}{\emph{PACIS}}.
  \bibinfo{pages}{141}.
\newblock


\bibitem[Hovy et~al\mbox{.}(2014)]%
        {hovy2014experiments}
\bibfield{author}{\bibinfo{person}{Dirk Hovy}, \bibinfo{person}{Barbara Plank},
  {and} \bibinfo{person}{Anders S{\o}gaard}.} \bibinfo{year}{2014}\natexlab{}.
\newblock \showarticletitle{Experiments with crowdsourced re-annotation of a
  POS tagging data set}. In \bibinfo{booktitle}{\emph{Proceedings of the 52nd
  Annual Meeting of the Association for Computational Linguistics (Volume 2:
  Short Papers)}}. \bibinfo{pages}{377--382}.
\newblock


\bibitem[Hube et~al\mbox{.}(2019)]%
        {hube2019understanding}
\bibfield{author}{\bibinfo{person}{Christoph Hube}, \bibinfo{person}{Besnik
  Fetahu}, {and} \bibinfo{person}{Ujwal Gadiraju}.}
  \bibinfo{year}{2019}\natexlab{}.
\newblock \showarticletitle{Understanding and mitigating worker biases in the
  crowdsourced collection of subjective judgments}. In
  \bibinfo{booktitle}{\emph{Proceedings of the 2019 CHI Conference on Human
  Factors in Computing Systems}}. \bibinfo{publisher}{ACM},
  \bibinfo{pages}{1--12}.
\newblock
\urldef\tempurl%
\url{https://doi.org/10.1145/3290605.3300637}
\showDOI{\tempurl}


\bibitem[I{\~n}iguez et~al\mbox{.}(2022)]%
        {iniguez2022dynamics}
\bibfield{author}{\bibinfo{person}{Gerardo I{\~n}iguez},
  \bibinfo{person}{Carlos Pineda}, \bibinfo{person}{Carlos Gershenson}, {and}
  \bibinfo{person}{Albert-L{\'a}szl{\'o} Barab{\'a}si}.}
  \bibinfo{year}{2022}\natexlab{}.
\newblock \showarticletitle{Dynamics of ranking}.
\newblock \bibinfo{journal}{\emph{Nature Communications}} \bibinfo{volume}{13},
  \bibinfo{number}{1} (\bibinfo{year}{2022}), \bibinfo{pages}{1--7}.
\newblock
\urldef\tempurl%
\url{https://doi.org/10.1038/s41467-022-29256-x}
\showDOI{\tempurl}


\bibitem[Jang et~al\mbox{.}(2022)]%
        {jang2022decreasing}
\bibfield{author}{\bibinfo{person}{Ikbeom Jang}, \bibinfo{person}{Garrison
  Danley}, \bibinfo{person}{Ken Chang}, {and} \bibinfo{person}{Jayashree
  Kalpathy-Cramer}.} \bibinfo{year}{2022}\natexlab{}.
\newblock \showarticletitle{Decreasing Annotation Burden of Pairwise
  Comparisons with Human-in-the-Loop Sorting: Application in Medical Image
  Artifact Rating}.
\newblock \bibinfo{journal}{\emph{arXiv preprint}} (\bibinfo{year}{2022}).
\newblock
\urldef\tempurl%
\url{https://doi.org/10.48550/arXiv.2202.04823}
\showDOI{\tempurl}


\bibitem[Kairam and Heer(2016)]%
        {kairam2016parting}
\bibfield{author}{\bibinfo{person}{Sanjay Kairam} {and}
  \bibinfo{person}{Jeffrey Heer}.} \bibinfo{year}{2016}\natexlab{}.
\newblock \showarticletitle{Parting crowds: Characterizing divergent
  interpretations in crowdsourced annotation tasks}. In
  \bibinfo{booktitle}{\emph{Proceedings of the 19th ACM Conference on
  Computer-Supported Cooperative Work \& Social Computing}}.
  \bibinfo{pages}{1637--1648}.
\newblock
\urldef\tempurl%
\url{https://doi.org/10.1145/2818048.2820016}
\showDOI{\tempurl}


\bibitem[Kane et~al\mbox{.}(2020)]%
        {kane2020nubia}
\bibfield{author}{\bibinfo{person}{Hassan Kane},
  \bibinfo{person}{Muhammed~Yusuf Kocyigit}, \bibinfo{person}{Ali Abdalla},
  \bibinfo{person}{Pelkins Ajanoh}, {and} \bibinfo{person}{Mohamed Coulibali}.}
  \bibinfo{year}{2020}\natexlab{}.
\newblock \showarticletitle{NUBIA: NeUral based interchangeability assessor for
  text generation}.
\newblock \bibinfo{journal}{\emph{arXiv preprint arXiv:2004.14667}}
  (\bibinfo{year}{2020}).
\newblock


\bibitem[Kittur et~al\mbox{.}(2013)]%
        {kittur2013future}
\bibfield{author}{\bibinfo{person}{Aniket Kittur}, \bibinfo{person}{Jeffrey~V
  Nickerson}, \bibinfo{person}{Michael Bernstein}, \bibinfo{person}{Elizabeth
  Gerber}, \bibinfo{person}{Aaron Shaw}, \bibinfo{person}{John Zimmerman},
  \bibinfo{person}{Matt Lease}, {and} \bibinfo{person}{John Horton}.}
  \bibinfo{year}{2013}\natexlab{}.
\newblock \showarticletitle{The future of crowd work}. In
  \bibinfo{booktitle}{\emph{Proceedings of the 2013 conference on Computer
  supported cooperative work}}. \bibinfo{publisher}{{ACM}},
  \bibinfo{pages}{1301--1318}.
\newblock
\urldef\tempurl%
\url{https://doi.org/10.1145/2441776.2441923}
\showDOI{\tempurl}


\bibitem[Knowlton(1966)]%
        {knowlton1966definition}
\bibfield{author}{\bibinfo{person}{James~Q Knowlton}.}
  \bibinfo{year}{1966}\natexlab{}.
\newblock \showarticletitle{On the definition of “picture”}.
\newblock \bibinfo{journal}{\emph{AV communication review}}
  \bibinfo{volume}{14}, \bibinfo{number}{2} (\bibinfo{year}{1966}),
  \bibinfo{pages}{157--183}.
\newblock
\urldef\tempurl%
\url{https://doi.org/10.1007/BF02769550}
\showDOI{\tempurl}


\bibitem[Kovashka and Grauman(2013)]%
        {kovashka2013attribute}
\bibfield{author}{\bibinfo{person}{Adriana Kovashka} {and}
  \bibinfo{person}{Kristen Grauman}.} \bibinfo{year}{2013}\natexlab{}.
\newblock \showarticletitle{Attribute adaptation for personalized image
  search}. In \bibinfo{booktitle}{\emph{Proceedings of the IEEE International
  Conference on Computer Vision}}. \bibinfo{publisher}{{IEEE} Computer
  Society}, \bibinfo{pages}{3432--3439}.
\newblock
\urldef\tempurl%
\url{https://doi.org/10.1109/ICCV.2013.426}
\showDOI{\tempurl}


\bibitem[Kuang et~al\mbox{.}(2020)]%
        {kuang2020spam}
\bibfield{author}{\bibinfo{person}{Li Kuang}, \bibinfo{person}{Huan Zhang},
  \bibinfo{person}{Ruyi Shi}, \bibinfo{person}{Zhifang Liao}, {and}
  \bibinfo{person}{Xiaoxian Yang}.} \bibinfo{year}{2020}\natexlab{}.
\newblock \showarticletitle{A spam worker detection approach based on
  heterogeneous network embedding in crowdsourcing platforms}.
\newblock \bibinfo{journal}{\emph{Computer Networks}}  \bibinfo{volume}{183}
  (\bibinfo{year}{2020}), \bibinfo{pages}{107587}.
\newblock


\bibitem[Li et~al\mbox{.}(2020)]%
        {li2020towards}
\bibfield{author}{\bibinfo{person}{Yanying Li}, \bibinfo{person}{Haipei Sun},
  {and} \bibinfo{person}{Wendy~Hui Wang}.} \bibinfo{year}{2020}\natexlab{}.
\newblock \showarticletitle{Towards fair truth discovery from biased
  crowdsourced answers}. In \bibinfo{booktitle}{\emph{Proceedings of the 26th
  ACM SIGKDD International Conference on Knowledge Discovery \& Data Mining}}.
  \bibinfo{publisher}{{ACM}}, \bibinfo{pages}{599--607}.
\newblock
\urldef\tempurl%
\url{https://doi.org/10.1145/3394486.3403102}
\showDOI{\tempurl}


\bibitem[Logeswaran et~al\mbox{.}(2018)]%
        {logeswaran2018content}
\bibfield{author}{\bibinfo{person}{Lajanugen Logeswaran},
  \bibinfo{person}{Honglak Lee}, {and} \bibinfo{person}{Samy Bengio}.}
  \bibinfo{year}{2018}\natexlab{}.
\newblock \showarticletitle{Content preserving text generation with attribute
  controls}.
\newblock \bibinfo{journal}{\emph{Advances in Neural Information Processing
  Systems}}  \bibinfo{volume}{31} (\bibinfo{year}{2018}).
\newblock


\bibitem[Marchenko et~al\mbox{.}(2020)]%
        {marchenko2020improving}
\bibfield{author}{\bibinfo{person}{OO Marchenko}, \bibinfo{person}{OS
  Radyvonenko}, \bibinfo{person}{TS Ignatova}, \bibinfo{person}{PV Titarchuk},
  {and} \bibinfo{person}{DV Zhelezniakov}.} \bibinfo{year}{2020}\natexlab{}.
\newblock \showarticletitle{Improving text generation through introducing
  coherence metrics}.
\newblock \bibinfo{journal}{\emph{Cybernetics and Systems Analysis}}
  \bibinfo{volume}{56}, \bibinfo{number}{1} (\bibinfo{year}{2020}),
  \bibinfo{pages}{13--21}.
\newblock


\bibitem[Menke and Martinez(2008)]%
        {menke2008bradley}
\bibfield{author}{\bibinfo{person}{Joshua~E Menke} {and}
  \bibinfo{person}{Tony~R Martinez}.} \bibinfo{year}{2008}\natexlab{}.
\newblock \showarticletitle{A Bradley--Terry artificial neural network model
  for individual ratings in group competitions}.
\newblock \bibinfo{journal}{\emph{Neural computing and Applications}}
  \bibinfo{volume}{17}, \bibinfo{number}{2} (\bibinfo{year}{2008}),
  \bibinfo{pages}{175--186}.
\newblock


\bibitem[Mohanty et~al\mbox{.}(2019)]%
        {mohanty2019second}
\bibfield{author}{\bibinfo{person}{Vikram Mohanty}, \bibinfo{person}{Kareem
  Abdol-Hamid}, \bibinfo{person}{Courtney Ebersohl}, {and}
  \bibinfo{person}{Kurt Luther}.} \bibinfo{year}{2019}\natexlab{}.
\newblock \showarticletitle{Second opinion: Supporting last-mile person
  identification with crowdsourcing and face recognition}. In
  \bibinfo{booktitle}{\emph{Proceedings of the AAAI Conference on Human
  Computation and Crowdsourcing}}, Vol.~\bibinfo{volume}{7}.
  \bibinfo{pages}{86--96}.
\newblock
\urldef\tempurl%
\url{https://doi.org/10.1609/hcomp.v7i1.5272}
\showDOI{\tempurl}


\bibitem[Mohanty et~al\mbox{.}(2018)]%
        {mohanty20181}
\bibfield{author}{\bibinfo{person}{Vikram Mohanty}, \bibinfo{person}{David
  Thames}, {and} \bibinfo{person}{Kurt Luther}.}
  \bibinfo{year}{2018}\natexlab{}.
\newblock \showarticletitle{Are 1,000 features worth a picture? combining
  crowdsourcing and face recognition to identify civil war soldiers}. In
  \bibinfo{booktitle}{\emph{AAAI Conference on Human Computation and
  Crowdsourcing (HCOMP 2018)}}.
\newblock
\urldef\tempurl%
\url{https://par.nsf.gov/biblio/10081893}
\showURL{%
\tempurl}


\bibitem[Ogden and Richards(1925)]%
        {ogden1925meaning}
\bibfield{author}{\bibinfo{person}{Charles~Kay Ogden} {and}
  \bibinfo{person}{Ivor~Armstrong Richards}.} \bibinfo{year}{1925}\natexlab{}.
\newblock \bibinfo{booktitle}{\emph{The Meaning of Meaning: A Study of the
  Influence of Language upon Thought and of the Science of Symbolism}}.
  Vol.~\bibinfo{volume}{29}.
\newblock \bibinfo{publisher}{Harcourt, Brace}.
\newblock


\bibitem[Parikh and Grauman(2011)]%
        {parikh2011relative}
\bibfield{author}{\bibinfo{person}{Devi Parikh} {and} \bibinfo{person}{Kristen
  Grauman}.} \bibinfo{year}{2011}\natexlab{}.
\newblock \showarticletitle{Relative attributes}. In
  \bibinfo{booktitle}{\emph{2011 International Conference on Computer Vision}}.
  IEEE, \bibinfo{publisher}{{IEEE} Computer Society},
  \bibinfo{pages}{503--510}.
\newblock
\urldef\tempurl%
\url{https://doi.org/10.1109/ICCV.2011.6126281}
\showDOI{\tempurl}


\bibitem[Porter et~al\mbox{.}(2020)]%
        {porter2020enhancing}
\bibfield{author}{\bibinfo{person}{Nathaniel~D Porter},
  \bibinfo{person}{Ashton~M Verdery}, {and} \bibinfo{person}{S~Michael
  Gaddis}.} \bibinfo{year}{2020}\natexlab{}.
\newblock \showarticletitle{Enhancing big data in the social sciences with
  crowdsourcing: Data augmentation practices, techniques, and opportunities}.
\newblock \bibinfo{journal}{\emph{PloS one}} \bibinfo{volume}{15},
  \bibinfo{number}{6} (\bibinfo{year}{2020}), \bibinfo{pages}{e0233154}.
\newblock


\bibitem[Ribeiro et~al\mbox{.}(2011)]%
        {ribeiro2011crowdsourcing}
\bibfield{author}{\bibinfo{person}{Fl{\'a}vio Ribeiro}, \bibinfo{person}{Dinei
  Florencio}, {and} \bibinfo{person}{V{\'\i}tor Nascimento}.}
  \bibinfo{year}{2011}\natexlab{}.
\newblock \showarticletitle{Crowdsourcing subjective image quality evaluation}.
  In \bibinfo{booktitle}{\emph{2011 18th IEEE International Conference on Image
  Processing}}. IEEE, \bibinfo{publisher}{IEEE}, \bibinfo{pages}{3097--3100}.
\newblock
\urldef\tempurl%
\url{https://doi.org/10.1109/ICIP.2011.6116320}
\showDOI{\tempurl}


\bibitem[Salimans et~al\mbox{.}(2016)]%
        {salimans2016improved}
\bibfield{author}{\bibinfo{person}{Tim Salimans}, \bibinfo{person}{Ian
  Goodfellow}, \bibinfo{person}{Wojciech Zaremba}, \bibinfo{person}{Vicki
  Cheung}, \bibinfo{person}{Alec Radford}, {and} \bibinfo{person}{Xi Chen}.}
  \bibinfo{year}{2016}\natexlab{}.
\newblock \showarticletitle{Improved techniques for training gans}.
\newblock \bibinfo{journal}{\emph{Advances in neural information processing
  systems}}  \bibinfo{volume}{29} (\bibinfo{year}{2016}).
\newblock


\bibitem[Salminen et~al\mbox{.}(2019)]%
        {salminen2019online}
\bibfield{author}{\bibinfo{person}{Joni Salminen}, \bibinfo{person}{Hind
  Almerekhi}, \bibinfo{person}{Ahmed~Mohamed Kamel}, \bibinfo{person}{Soon-gyo
  Jung}, {and} \bibinfo{person}{Bernard~J Jansen}.}
  \bibinfo{year}{2019}\natexlab{}.
\newblock \showarticletitle{Online hate ratings vary by extremes: A statistical
  analysis}. In \bibinfo{booktitle}{\emph{Proceedings of the 2019 Conference on
  Human Information Interaction and Retrieval}}. \bibinfo{pages}{213--217}.
\newblock


\bibitem[Salminen et~al\mbox{.}(2021)]%
        {salminen2021problem}
\bibfield{author}{\bibinfo{person}{Joni Salminen},
  \bibinfo{person}{Ahmed~Mohamed Kamel}, \bibinfo{person}{Soon-Gyo Jung}, {and}
  \bibinfo{person}{Bernard Jansen}.} \bibinfo{year}{2021}\natexlab{}.
\newblock \showarticletitle{The Problem of Majority Voting in Crowdsourcing
  with Binary Classes}. In \bibinfo{booktitle}{\emph{Proceedings of 19th
  European Conference on Computer-Supported Cooperative Work}}. European
  Society for Socially Embedded Technologies (EUSSET).
\newblock


\bibitem[Salminen et~al\mbox{.}(2018b)]%
        {salminen2018online}
\bibfield{author}{\bibinfo{person}{Joni Salminen}, \bibinfo{person}{Fabio
  Veronesi}, \bibinfo{person}{Hind Almerekhi}, \bibinfo{person}{Soon-Gvo Jung},
  {and} \bibinfo{person}{Bernard~J Jansen}.} \bibinfo{year}{2018}\natexlab{b}.
\newblock \showarticletitle{Online hate interpretation varies by country, but
  more by individual: A statistical analysis using crowdsourced ratings}. In
  \bibinfo{booktitle}{\emph{2018 Fifth International Conference on Social
  Networks Analysis, Management and Security (SNAMS)}}. IEEE,
  \bibinfo{pages}{88--94}.
\newblock
\urldef\tempurl%
\url{https://doi.org/10.1109/SNAMS.2018.8554954}
\showDOI{\tempurl}


\bibitem[Salminen et~al\mbox{.}(2018a)]%
        {salminen2018inter}
\bibfield{author}{\bibinfo{person}{Joni~O Salminen}, \bibinfo{person}{Hind~A
  Al-Merekhi}, \bibinfo{person}{Partha Dey}, {and} \bibinfo{person}{Bernard~J
  Jansen}.} \bibinfo{year}{2018}\natexlab{a}.
\newblock \showarticletitle{Inter-rater agreement for social computing
  studies}. In \bibinfo{booktitle}{\emph{2018 Fifth International Conference on
  Social Networks Analysis, Management and Security (SNAMS)}}. IEEE,
  \bibinfo{pages}{80--87}.
\newblock


\bibitem[Snow et~al\mbox{.}(2008)]%
        {snow2008cheap}
\bibfield{author}{\bibinfo{person}{Rion Snow}, \bibinfo{person}{Brendan
  O’connor}, \bibinfo{person}{Dan Jurafsky}, {and} \bibinfo{person}{Andrew~Y
  Ng}.} \bibinfo{year}{2008}\natexlab{}.
\newblock \showarticletitle{Cheap and fast--but is it good? evaluating
  non-expert annotations for natural language tasks}. In
  \bibinfo{booktitle}{\emph{Proceedings of the 2008 conference on empirical
  methods in natural language processing}}. \bibinfo{pages}{254--263}.
\newblock


\bibitem[Sobieraj and Berry(2011)]%
        {sobieraj2011incivility}
\bibfield{author}{\bibinfo{person}{Sarah Sobieraj} {and}
  \bibinfo{person}{Jeffrey~M Berry}.} \bibinfo{year}{2011}\natexlab{}.
\newblock \showarticletitle{From incivility to outrage: Political discourse in
  blogs, talk radio, and cable news}.
\newblock \bibinfo{journal}{\emph{Political Communication}}
  \bibinfo{volume}{28}, \bibinfo{number}{1} (\bibinfo{year}{2011}),
  \bibinfo{pages}{19--41}.
\newblock


\bibitem[Stern(1992)]%
        {stern1992all}
\bibfield{author}{\bibinfo{person}{Hal Stern}.}
  \bibinfo{year}{1992}\natexlab{}.
\newblock \showarticletitle{Are all linear paired comparison models empirically
  equivalent?}
\newblock \bibinfo{journal}{\emph{Mathematical Social Sciences}}
  \bibinfo{volume}{23}, \bibinfo{number}{1} (\bibinfo{year}{1992}),
  \bibinfo{pages}{103--117}.
\newblock
\urldef\tempurl%
\url{https://doi.org/10.1016/0165-4896(92)90040-C}
\showDOI{\tempurl}


\bibitem[Sumner et~al\mbox{.}(2020)]%
        {sumner2020crowdsourcing}
\bibfield{author}{\bibinfo{person}{Jane~Lawrence Sumner},
  \bibinfo{person}{Emily~M Farris}, {and} \bibinfo{person}{Mirya~R Holman}.}
  \bibinfo{year}{2020}\natexlab{}.
\newblock \showarticletitle{Crowdsourcing reliable local data}.
\newblock \bibinfo{journal}{\emph{Political Analysis}} \bibinfo{volume}{28},
  \bibinfo{number}{2} (\bibinfo{year}{2020}), \bibinfo{pages}{244--262}.
\newblock


\bibitem[Sunahase et~al\mbox{.}(2017)]%
        {sunahase2017pairwise}
\bibfield{author}{\bibinfo{person}{Takeru Sunahase}, \bibinfo{person}{Yukino
  Baba}, {and} \bibinfo{person}{Hisashi Kashima}.}
  \bibinfo{year}{2017}\natexlab{}.
\newblock \showarticletitle{Pairwise hits: Quality estimation from pairwise
  comparisons in creator-evaluator crowdsourcing process}. In
  \bibinfo{booktitle}{\emph{Proceedings of the AAAI Conference on Artificial
  Intelligence}}, Vol.~\bibinfo{volume}{31}.
\newblock


\bibitem[Tao et~al\mbox{.}(2020)]%
        {tao2020label}
\bibfield{author}{\bibinfo{person}{Fangna Tao}, \bibinfo{person}{Liangxiao
  Jiang}, {and} \bibinfo{person}{Chaoqun Li}.} \bibinfo{year}{2020}\natexlab{}.
\newblock \showarticletitle{Label similarity-based weighted soft majority
  voting and pairing for crowdsourcing}.
\newblock \bibinfo{journal}{\emph{Knowledge and Information Systems}}
  \bibinfo{volume}{62}, \bibinfo{number}{7} (\bibinfo{year}{2020}),
  \bibinfo{pages}{2521--2538}.
\newblock


\bibitem[Uma et~al\mbox{.}(2021)]%
        {uma2021learning}
\bibfield{author}{\bibinfo{person}{Alexandra~N Uma}, \bibinfo{person}{Tommaso
  Fornaciari}, \bibinfo{person}{Dirk Hovy}, \bibinfo{person}{Silviu Paun},
  \bibinfo{person}{Barbara Plank}, {and} \bibinfo{person}{Massimo Poesio}.}
  \bibinfo{year}{2021}\natexlab{}.
\newblock \showarticletitle{Learning from disagreement: A survey}.
\newblock \bibinfo{journal}{\emph{Journal of Artificial Intelligence Research}}
   \bibinfo{volume}{72} (\bibinfo{year}{2021}), \bibinfo{pages}{1385--1470}.
\newblock
\urldef\tempurl%
\url{https://doi.org/10.1613/jair.1.12752}
\showDOI{\tempurl}


\bibitem[Vuurens et~al\mbox{.}(2011)]%
        {vuurens2011much}
\bibfield{author}{\bibinfo{person}{Jeroen Vuurens}, \bibinfo{person}{Arjen~P de
  Vries}, {and} \bibinfo{person}{Carsten Eickhoff}.}
  \bibinfo{year}{2011}\natexlab{}.
\newblock \showarticletitle{How much spam can you take? an analysis of
  crowdsourcing results to increase accuracy}. In
  \bibinfo{booktitle}{\emph{Proc. ACM SIGIR Workshop on Crowdsourcing for
  Information Retrieval (CIR’11)}}. \bibinfo{publisher}{{ACM}},
  \bibinfo{pages}{21--26}.
\newblock


\bibitem[Wallace et~al\mbox{.}(2022)]%
        {wallace2022debiased}
\bibfield{author}{\bibinfo{person}{Shaun Wallace}, \bibinfo{person}{Tianyuan
  Cai}, \bibinfo{person}{Brendan Le}, {and} \bibinfo{person}{Luis~A Leiva}.}
  \bibinfo{year}{2022}\natexlab{}.
\newblock \showarticletitle{Debiased Label Aggregation for Subjective
  Crowdsourcing Tasks}. In \bibinfo{booktitle}{\emph{CHI Conference on Human
  Factors in Computing Systems Extended Abstracts}}. \bibinfo{pages}{1--8}.
\newblock
\urldef\tempurl%
\url{https://doi.org/10.1145/3491101.3519614}
\showDOI{\tempurl}


\bibitem[Wallace et~al\mbox{.}(2020)]%
        {wallace2020sketchy}
\bibfield{author}{\bibinfo{person}{Shaun Wallace}, \bibinfo{person}{Brendan
  Le}, \bibinfo{person}{Luis~A Leiva}, \bibinfo{person}{Aman Haq},
  \bibinfo{person}{Ari Kintisch}, \bibinfo{person}{Gabrielle Bufrem},
  \bibinfo{person}{Linda Chang}, {and} \bibinfo{person}{Jeff Huang}.}
  \bibinfo{year}{2020}\natexlab{}.
\newblock \showarticletitle{Sketchy: Drawing inspiration from the crowd}.
\newblock \bibinfo{journal}{\emph{Proceedings of the ACM on Human-Computer
  Interaction}} \bibinfo{volume}{4}, \bibinfo{number}{CSCW2}
  (\bibinfo{year}{2020}), \bibinfo{pages}{1--27}.
\newblock
\urldef\tempurl%
\url{https://doi.org/10.1145/3415243}
\showDOI{\tempurl}


\bibitem[Wu et~al\mbox{.}(2021)]%
        {wu2021toward}
\bibfield{author}{\bibinfo{person}{Jiele Wu}, \bibinfo{person}{Chau-Wai Wong},
  \bibinfo{person}{Xinyan Zhao}, {and} \bibinfo{person}{Xianpeng Liu}.}
  \bibinfo{year}{2021}\natexlab{}.
\newblock \showarticletitle{Toward effective automated content analysis via
  crowdsourcing}. In \bibinfo{booktitle}{\emph{2021 IEEE International
  Conference on Multimedia and Expo (ICME)}}. IEEE,
  \bibinfo{publisher}{{IEEE}}, \bibinfo{pages}{1--6}.
\newblock
\urldef\tempurl%
\url{https://doi.org/10.1109/ICME51207.2021.9428220}
\showDOI{\tempurl}


\bibitem[Ye and Doermann(2013)]%
        {ye2013combining}
\bibfield{author}{\bibinfo{person}{Peng Ye} {and} \bibinfo{person}{David
  Doermann}.} \bibinfo{year}{2013}\natexlab{}.
\newblock \showarticletitle{Combining preference and absolute judgements in a
  crowd-sourced setting}. In \bibinfo{booktitle}{\emph{ICML Workshop}}.
  \bibinfo{pages}{1--7}.
\newblock


\bibitem[Zheng et~al\mbox{.}(2017)]%
        {zheng2017truth}
\bibfield{author}{\bibinfo{person}{Yudian Zheng}, \bibinfo{person}{Guoliang
  Li}, \bibinfo{person}{Yuanbing Li}, \bibinfo{person}{Caihua Shan}, {and}
  \bibinfo{person}{Reynold Cheng}.} \bibinfo{year}{2017}\natexlab{}.
\newblock \showarticletitle{Truth inference in crowdsourcing: Is the problem
  solved?}
\newblock \bibinfo{journal}{\emph{Proceedings of the VLDB Endowment}}
  \bibinfo{volume}{10}, \bibinfo{number}{5} (\bibinfo{year}{2017}),
  \bibinfo{pages}{541--552}.
\newblock
\urldef\tempurl%
\url{https://doi.org/10.14778/3055540.3055547}
\showDOI{\tempurl}


\bibitem[Zou et~al\mbox{.}(2015)]%
        {zou2015crowdsourcing}
\bibfield{author}{\bibinfo{person}{James Zou}, \bibinfo{person}{Kamalika
  Chaudhuri}, {and} \bibinfo{person}{Adam Kalai}.}
  \bibinfo{year}{2015}\natexlab{}.
\newblock \showarticletitle{Crowdsourcing feature discovery via adaptively
  chosen comparisons}. In \bibinfo{booktitle}{\emph{Proceedings of the AAAI
  Conference on Human Computation and Crowdsourcing}},
  Vol.~\bibinfo{volume}{3}. \bibinfo{pages}{198--205}.
\newblock
\urldef\tempurl%
\url{https://doi.org/10.1609/hcomp.v3i1.13231}
\showDOI{\tempurl}


\end{thebibliography}

\appendix
\section{Appendix: Using the comparison method in the real-world}\label{appendix:application}
In this paper, we discussed merits and properties of the comparison method for labelling subjective constructs. In this section we will provide a brief summary of our recommended approach as a step-by-step guide, as well as briefly introduce methods of dealing with spamming or inattentive raters.

Let us assume that you have a set of items and an intended subjective construct in mind. This approach is designed for subjective qualities where one can say A has more (or less) of this quality than B, such as the toxicity of tweets or realisticness of an AI generated image.

Depending on the question at hand, you need a representative sample of items to be compared. If you intend to compare relative toxicity of all English language tweets on the subject of climate change compared to a baseline month, you need a representative sample of English climate change tweets. Similarly, attention has to be paid on selecting the rater population, as the rating and labels derived from this method is representative of the opinions and the viewpoint of the raters.

The next step is to run an exhaustive pilot, similar to Fig.~\ref{fig:scaling}(d), for a smaller subset of your sample. For example, for gathering a larger dataset of controversial and non-controversial tweets for training machine-learning models or to study the evolution of controversiality of climate tweets across time, researchers might consider running a pilot experiment similar to what was described in Sec.~\ref{sec:experiments}. Select a sub-sample of your items and perform all possible comparisons between pairs of items. At this step you can elect to perform each possible comparison multiple times by different raters, this also allows you to compare their answers using existing rater agreement statistics.

Plot the result of this pilot study in the same manner as Fig.~\ref{fig:scaling}(d). The scaling trajectory of the pilot experiment would allow you to select the appropriate compromise between labelling accuracy and number of items labelled, based on the budget allocated for the study. Let us say you require that the items in your final output have a labelling accuracy, as indicated by $f_1$ score, of at least 0.9. The x-axis value where the scaling trajectory intersects $y = 0.9$ predicts a value for $\frac{n_\text{comparisons}}{N \log N}$ where $N$ stands for the number of items and $n_\text{comparisons}$ for the number of comparisons. If your budget allows for a certain maximum number of crowdsourcing tasks, solving for $N$ gives you the number of items that you can label. For the experiment performed in Sec.~\ref{sec:experiments}, to get a $f_1$ score of 0.8 for 1000 unique conversations for this specific task, you can expect to need close to 16000 comparisons based on the scaling plot Fig.~\ref{fig:scaling}(d).

Note that the trajectory of the scaling plot is subject to the properties of the tasks at hand, meaning that a different research question or a significantly different population of raters might require its own specific pilot study and that task subjectivity parameters such as ambiguity in comparison and perception can affect the number of required pairwise comparisons and the result of the Elo system for a specific value of accuracy. As we do not propose a direct method for measuring these properties, the most reliable approach of ensuring the efficacy of the full-scale experiment is to determine the number of required comparisons for $N$ items to achieve high-quality results by conducting pilot experiments.  

After determining the number of items $N$ in the full-scale experiment based on the budget and the pilot experiment results, select a sample of size $N$ from the items and perform random comparisons between pairs of items. Aggregating the results through the Elo rating system gives you a rating for each item, indicating the strength of subjective construct in that item as construed by the average rater. This rating can be translated into binary classification labels.

For both the pilot and the main study, we recommend cycling through all comparisons in random order for multiple epochs, until Elo ratings of each item arrives at a stable value. This can be viewed similar to the practice of cycling through the training data multiple times when training a deep neural network model, which here helps ease the negative effects of selecting a small value for parameter $k$ of the Elo method.

If your research question depends on comparing one population of items to another, let us say comparing toxicity of English climate change tweets to background level of toxicity of English in each month, you can draw separate samples for each population and calculating ratings separately for the baseline group and benchmark group. Subsequently, you can ``anchor'' the ratings from the baseline group to the benchmark group by, e.g., selecting the 5 items closest to the average in each month in the baseline group, performing comparisons using a binary-search algorithm for each of the 5 baseline tweets against sorted benchmark tweets belonging to the same month. This would give you an approximate ranking (and rating) for each of 5 baseline tweets among the benchmark tweets of the same month that can be used for labelling items in the benchmark set.

\subsection{Detecting spammers}\label{appendix:spam}
While, as we showed in Sec.~\ref{res:additional-results}, the comparison method overall is more robust toward inattentive raters or spammers compared to the majority-vote method, it might be necessary to have the ability to detect flagrant instances of spamming. While research in this field is still in its preliminary stage, we suggest multiple mitigation approaches. The most simple approach relies on the correlation between the difference in rating $R_A - R_B$ and the selected outcome $S_A$ in every comparison performed by one worker. We expect an attentive rater who has performed a sufficiently large number of comparisons to show a significant positive correlation, and the absence thereof or a negative correlation can be interpreted as a sign of possible spamming. Another approach is based on the probability of the selected outcome for each comparison performed by one rater. Equation.~\eqref{eq:elo} can be used to calculate a probability for the selected outcome of a comparison by a rater. Consistently selecting low-probability outcomes, as can be indicated by median selection probability, can signify possible spamming.

In all of the above-mentioned methods, when implementing these approaches to investigate a specific rater, it is advisable to calculate ratings and rankings based on the contributions of all raters except the rater being investigated.

Reliability of these approaches can be improved by having a larger set of comparisons performed by each rater, though this comes at a cost of reducing diversity among the raters. It is also possible to perform attention-check comparisons, consisting of items known to have large rating differences based on the pilot results. The results of these comparisons, when combined with each rater's comparisons can greatly bolster the previously mentioned spammer-detection approaches. Implementation of such an approach can be akin to the one from \citet{hettiachchi2020augmenting}, where their crowdsourced kinship verification had a number of categorically verifiable kinship queries, specifically in the form of asking whether a cartoon character looks like or is related to a human.

\subsection{Elo rating system reference implementation}
Our implementation of the Elo rating system is available as a Python package. This package can be installed using the command \anon[Redacted pip install command]{\texttt{python -m pip install -U elo-rating}}. The documentation as well as the source code can be found at \anon[Redacted documentation URL]{\url{https://github.com/hastinarimanzadeh/elo-rating}}. Listing~\ref{lst:elo-example} displays common usage pattern as a simple code snippet.

\begin{lstlisting}[caption={Example use-case of the Python Elo rating system implementation in aggregating comparisons.},label={lst:elo-example},language=Python,float,frame=single]
from elo_rating import Elo

# create an elo class, denom referes to 1/(logistic growth rate),
#   e.g. the number 400 in Eq. 2
e = Elo(denom=400)

# add comparisons one by one
e.add_match("p1", "p2", 1.0, k=30, default_rating=1400)

# or many at the same time
e.add_matches([("p1", "p2", 1.0), ("p2", "p1", 0.5)],
w    k=30, default_rating=1400)

# get a list of all seen items
e.items() # ['p1', 'p2']

# item ratings based on current comparisons
e.ratings()
# =>  {'p1': 1428.3358360702414, 'p2': 1371.6641639297586}

# item ratings based on current comparisons
e.rankings()
# => {'p1': 0, 'p2': 1}

e.ranking('p1') # 0
e.rating('p1')  # 1428.3358360702414

# probability of win for p1 in a hypothetical match with p2
e.expected_score("p1", "p2") # 0.5353606653429002
\end{lstlisting}

\end{document}